\documentclass{aa}

\usepackage{psfig}
\usepackage{aabib99}
\usepackage{graphicx} \usepackage{psfig}
\usepackage{rotating} \usepackage{times}

\newcommand{\zsun}{\ensuremath{Z_{\odot}}}
\newcommand{\hsun}{\ensuremath{\rm H_{\odot}}}
\newcommand{\fesun}{\ensuremath{\textrm{Fe}_{\odot}}}

\newcommand{\hb}{\ensuremath{{\textrm{H}}\beta}}
\newcommand{\hda}{\ensuremath{{\textrm{H}}\delta_{A}}}
\newcommand{\hdf}{\ensuremath{{\textrm{H}}\delta_{F}}}
\newcommand{\hga}{\ensuremath{{\textrm{H}}\gamma_{A}}}
\newcommand{\hgf}{\ensuremath{{\textrm{H}}\gamma_{F}}}
\newcommand{\mg}{\ensuremath{{\textrm{Mg}}_2}}
\newcommand{\mguno}{\ensuremath{{\textrm{Mg}}_1}}
\newcommand{\mgb}{\ensuremath{{\textrm{Mg}}\, b}}
\newcommand{\fe}{\ensuremath{\langle {\textrm{Fe}}\rangle}}
\newcommand{\teff}{\ensuremath{T_{\rm eff}}}
\newcommand{\afe}{\ensuremath{[\alpha/{\rm Fe}]}}
\newcommand{\feh}{\ensuremath[{{\rm Fe}/{\rm H}}]}
\newcommand{\zh}{\ensuremath[{{\rm Z}/{\rm H}}]}
\newcommand{\CNone}{\ensuremath{{\rm CN}_1}}
\newcommand{\CNtwo}{\ensuremath{{\rm CN}_2}}
\newcommand{\ferrofake}{\ensuremath{{\rm C}_2}4668}
\newcommand{\TiOone}{\ensuremath{{\rm TiO}_1}}
\newcommand{\TiOtwo}{\ensuremath{{\rm TiO}_2}}

\newcommand{\issp}{\ensuremath{I^{\rm SSP}}} 
\newcommand{\istar}{\ensuremath{I{\rm _i^*}}}
\newcommand{\istarj}{\ensuremath{I{\rm _j}^*}}
\newcommand{\istari}{\ensuremath{I{\rm _i}^*}}
\newcommand{\istarnoi}{\ensuremath{I{\rm ^*}}}

\newcommand{\Fcstar}{\ensuremath{F{\rm_{c,i}^*}}}
\newcommand{\fcstarj}{\ensuremath{f{\rm_{c,j}^*}}}

\newcommand{\gapprox}{\,\rlap{\lower 2.5pt 
\hbox{$\sim$}}\raise 1.5pt\hbox{$>$}\,}
\newcommand{\lapprox}{\,\rlap{\lower 2.5pt 
\hbox{$\sim$}}\raise 1.5pt\hbox{$<$}\,}

\begin{document}

   \titlerunning{Lick indices of Bulge stellar populations. II. Implications for SSP models and elliptical galaxies}

   \authorrunning{C. Maraston et al.}

   \title{Integrated Spectroscopy of Bulge Globular Clusters and Fields.\\ 
II. Implications for population synthesis models and elliptical galaxies}

     \author{C.~Maraston \inst{1}$^,$\inst{2}
	  \and
	  L.~Greggio \inst{2}$^,$\inst{3}
           \and 
	  A.~Renzini \inst{4}
	    \and
	  S.~Ortolani \inst{3}
	    \and
 	  R.~P.~Saglia \inst{2}
            \and
          T.~H.~Puzia \inst{2}
            \and 
         M.~Kissler-Patig \inst{4} 	  
               }

   \offprints{C. Maraston}

 \institute{Max-Planck-Institut f\"ur Extraterrestrische Physik,
              Giessenbachstra{\ss}e,
                D-85748 Garching b. M\"unchen, Germany\\
                email: maraston@mpe.mpg.de
	\and
      Universit\"ats-Sternwarte M\"unchen, Scheinerstr. 1, 
      D-81679 M\"unchen, Germany
        \and
     Osservatorio Astronomico di Padova, vicolo dell'Osservatorio, 
     Padova, Italy
	\and
     European Southern Observatory,
     Karl-Schwarzschild-Str.~2, 85748 Garching, Germany
  } 
  
   \date{Received ; accepted }

\abstract{An empirical calibration is presented for the synthetic Lick
indices (e.g.~\mg,~\fe,~\hb, etc.) of Simple Stellar Population (SSP)
models that for the first time extends up to solar metallicity.  This
is accomplished by means of a sample of Milky Way globular clusters
(GCs) whose metallicities range from $\sim \zsun/30$ to $Z\sim\zsun$,
thanks to the inclusion of several metal rich clusters belonging to
the Galactic bulge (e.g., NGC~6553 and NGC~6528).  This metallicity
range approaches the regime that is relevant for the interpretation of
the integrated spectra of elliptical galaxies. It is shown that the
spectra of both the globular clusters and the Galactic bulge follow
the same correlation between magnesium and iron indices that extends
to elliptical galaxies, showing weaker iron indices at given magnesium
indices with respect to the predictions of models that assume
solar-scaled abundances.  This similarity provides robust empirical
evidence for enhanced \afe~ratios in the stellar populations of
elliptical galaxies, since the globular clusters are independently
known to have enhanced \afe~ratios from spectroscopy of individual
stars.  We check the uniqueness of this $\alpha$-overabundance
solution by exploring the whole range of model ingredients and
parameters, i.e. fitting functions, stellar tracks, and the initial
mass function (IMF). We argue that the {\it standard} models (meant
for solar abundance ratios) succeed in reproducing the Mg-Fe
correlation at low metallicities ($\zh\lapprox-0.7$) because the
stellar templates used in the synthesis are Galactic halo stars that
actually are $\alpha$-enhanced. The same models, however, fail to
predict the observed Mg-Fe pattern at higher metallicities ($\zh
\gapprox-0.7$) (i.e., for bulge clusters and ellipticals alike)
because the high-metallicity templates are disk stars that are not
$\alpha$-enhanced.  We show that the new set of SSP models which
incorporates the dependence on the \afe~ ratio (Thomas, Maraston \&
Bender~2002) is able to reproduce the Mg and Fe indices of GCs at all
metallicities, with an $\alpha$-enhancement~$\afe=+0.3$, in agreement
with the available spectroscopic determinations. The \hb~index and the
higher-order Balmer indices are well calibrated, provided the
appropriate morphology of the Horizontal Branch is taken into
account. In particular, the Balmer line indices of the two metal rich
clusters NGC~6388 and NGC~6441, which are known to exhibit a tail of
warm Horizontal Branch stars, are well reproduced.  Finally, we note
that the Mg indices of very metal-poor ($\zh \lapprox-1.8$)
populations are dominated by the contribution of the lower Main
Sequence, hence are strongly affected by the present-day mass function
of individual globular clusters, which is known to vary from cluster
to cluster due to dynamical effects.
\keywords{Galaxy: globular clusters: general - galaxies: ellipticals
and lenticular, cD - galaxies: abundances - galaxies: evolution -
galaxies: formation} }

   \maketitle

\section{Introduction}
\label{intro}

Galactic spheroids, i.e., elliptical galaxies and the bulges of
spirals, include a major fraction (perhaps the majority) of the
stellar mass in the nearby universe (e.g., Fukugita et al.~1998). From
the uniformity of their fundamental properties at low as well as high
redshift ($z\lapprox 1$) it has been inferred that the bulk of stars
in spheroids must be very old, likely to have formed at $z>2-3$ (see
Renzini~1999 for a comprehensive review; see also
Peebles~2002). Moreover, the space density of passively evolving
galaxies at $z\sim 1$ is consistent with the bulk of massive spheroids
being already in place and 2-3 Gyr old at this early epoch (Cimatti et
al. 2002a,b). Current renditions of hierarchical galaxy formation in
CDM dominated universes have so far failed to predict these empirical
findings, favoring instead a late formation with major activity even
below $z \sim 1$. The culprit is probably in the ways in which star
formation and feedback processes have been parameterized and
implemented in the so-called semi-analytic models of galaxy formation
and evolution (e.g., Kauffmann \& Charlot 1998; Cole et al.  2001;
Somerville et al.~2001; Menci et al. 2002).

Given our poor understanding of the star formation and feedback
processes, the detailed study of the stellar populations in nearby
galaxies (the {\it fossil} records) can provide important clues on
their early formation phases, complementary to the direct observation
of very high-$z$ galaxies. Indeed, the ages and metallicities of the
stellar populations of a galaxy are useful constraints to its
formation mechanism.  However, the determination of absolute ages and
metallicities of composite stellar populations from their integrated
spectra is hampered by well known degeneracy effects (Faber~1972;
O'Connell~1976; Renzini~1986; Worthey~1994; Maraston \& Thomas~2000),
and a further complication arises when the abundances of major
chemical elements (iron, magnesium, oxygen, etc.) are considered, as
traced by narrow-band spectroscopic indices such as the so-called Lick
indices, \mg,~\fe,~\hb,~etc. (Burstein et al. 1984; Faber et al.~1985;
Worthey et al.~1994).  The application of this technique to
ellipticals revealed that the observed relation between magnesium and
iron indices (\mg~or \mgb~vs. \fe) disagrees with the predictions of
population synthesis models where the Mg/Fe ratio is assumed to be
solar. Observed magnesium indices at given iron are significantly {\it
stronger} than in the models (see Fig.~2 in Worthey et al.~1992). This
finding was confirmed by several subsequent studies for other samples
of ellipticals (Davies et al.~1993; Carollo \& Danziger~1994; Fisher
et al.~1995; Jorgensen~1997, Kuntschner \& Davies~1998; Mehlert et
al.~1998; Longhetti et al.~2000; Thomas et al.~2002a.)

If Lick indices of magnesium and iron trace the corresponding element
abundances, {\it and} the models that are meant for solar ratios of
these elements are correct, then the observed indices imply a
supersolar Mg/Fe ratio in ellipticals.  In turn, according to common
wisdom this implies short (${t\lapprox 1~\mathrm{Gyr}}$) star
formation timescales for the stellar populations of ellipticals (e.g.,
Matteucci~1994; Thomas et al.~1999). In fact, the
so-called $\alpha$-elements (i.e., O, Mg, Ca, Ti, and Si) are promptly
released by massive, short-living ($\lapprox 3\times 10^7$ yrs)
progenitors exploding as Type II supernovae, while most iron comes
from Type Ia supernovae, whose progenitors span evolutionary
timescales from over $\sim 3\times 10^7$ yrs to many Gyrs (e.g.,
Greggio \& Renzini 1983; Matteucci \& Greggio~1986;
Pagel~2001)). Therefore, a high $\alpha$-over-iron ratio (\afe)
implies that star formation ceased before the bulk of Type Ia
supernovae had the time to enrich with iron the interstellar medium
while this was still actively forming stars.  Such a short star
formation timescale appears to be at variance with the predictions of
current hierarchical models for the formation of elliptical galaxies
(Thomas~1999; Thomas \& Kauffmann~1999), which predict star formation
to continue for several Gyrs.  In conclusion, the Lick indices of
magnesium and iron appear to offer a unique opportunity to estimate
the timescale of star formation in galactic spheroids, hence to help
for a better understanding of the early formation phases of galaxies.

However, a caveat is in order over the above chain of arguments: how
well do Lick indices trace element abundances? Are we sure that
population synthesis models correctly predict the values of these
indices?  (for early discussions of these issues see Tripicco
\&~Bell~1995; Greggio~1997; Tantalo et al.~1998). Indeed, the
population synthesis models on the basis of which the magnesium
overabundance has been inferred were {\em not calibrated}, especially
in the metallicity range (solar and above) which is relevant to
elliptical galaxies. Hence, it could not be excluded that the
population synthesis models would underpredict the strength of the
magnesium indices, while the Mg/Fe ratio of ellipticals would actually
be solar.  By calibration of the indices we mean the comparison of
their synthetic values with the corresponding quantities measured on
objects for which the age and the detailed chemical composition -
total metallicity and element abundance ratios - are independently
known. For this comparison the best {\it stellar population templates}
are the Galactic globular clusters.  However, existing databases of
Lick indices of globular clusters (Burstein et al.~1984; Covino et
al.~1995) are restricted to the metal-poor objects of the Halo. The
most metal rich cluster in the Covino et al. sample is 47~Tuc ($\feh
\sim-0.7$) whose \mg~is $\sim 0.18$ mag, much less than found among
ellipticals, which span from \mg$\sim$ 0.2 to $\sim$ 0.4 mag.

Globular clusters that are more metal rich than 47~Tuc do actually
exist in the Galactic bulge, reaching $Z \sim \zsun$ for NGC~6553 and
NGC~6528 (Barbuy et al.~1999; Cohen et al. 1999). Cohen et al. (1998)
have measured some of the Lick indices for these two clusters, but did
so using the Burstein et al.~(1984) passbands to define the indices.
This does not allow a direct comparison with the theoretical models,
which are based on the passbands defined by Worthey et
al. (1994). Gregg~(1994) measures and analyses spectral indices for
several Milky Way GCs including metal-rich objects, among which
NGC~6528. Though similar, these spectral indices are not in the Lick
system. In Sect.~3 we show that in a model calibration it is crucial
that data and models are set up on the same system, because there are
sizable differences in the value of some of the indices, depending on
the adopted passbands. Therefore, we obtained optical spectra for a
sample of Bulge globular clusters with metallicities
$\feh\gapprox-0.5$ (including NGC~6528 and NGC~6553), plus some
metal-poor globular clusters in order to check the models on a wide
metallicity range. In fact, existing Lick indices of metal poor
globular clusters (e.g. Covino et al.~1995: Cohen et al. 1998) were
also measured in the Burstein et al.~(1984) system.  The results of
the measurement of the indices in the Lick/IDS system are reported in
an accompanying paper (Puzia et al.~2002, hereafter Paper~I).

For at least some of the program clusters the abundance ratios of
$\alpha$-elements to iron are known from high resolution spectroscopy
of individual stars in these clusters (Barbuy et al.~1999; Cohen et
al.~1999, Carretta et al.~2001; Coelho et al.~2001). While there is
certainly room for further improvements in the abundance
determinations, these studies indicate an overabundance $\afe \sim
0.2-0.3$~for these clusters.  Moreover, their age, determined from
color-magnitude diagrams, is virtually identical to the age of Halo
globular clusters, i.e., 12--13 Gyr (Ortolani et al.~1995; Rosenberg
et al.~1999; Feltzing \& Gilmore~2000, Zoccali et al.  2001).  Having
fairly accurate estimates for their basic parameters (age, \feh,
$\afe$, the measured Lick indices of Bulge globular clusters are used
to calibrate the population synthesis models and to test the
``magnesium overabundance'' solution for ellipticals.

The paper is organized as follows. A brief summary of the database is
presented in Sect.~2. In Sect.~3 a critical overview of existing
ambiguities in the definition of model and data metallicities, is
given.  The main results are presented in Sect.~4, where the Lick
indices of the Halo and Bulge clusters are compared to those of
elliptical galaxies. The following Sections explore in detail several
technical aspects of the population synthesis modeling (Sect.~5) and
the calibration of the models (Sect.~6). Readers mainly interested
in the implications of the results on galaxies can skip these
sections. Sect.~7 summarizes the conclusions.

\section[]{The Globular Cluster Data}
\label{data}

Optical spectra (3400~$<\lambda~<$~7500~\AA) have been obtained with the
Boller \& Chivens spectrograph at the ESO 1.5m telescope for a
sample of 12 globular clusters (GCs) mostly located in the Galactic
bulge and  for several (15) positions in the bulge field known as Baade's Window. 
The data acquisition, reduction and the resulting indices  in the Lick
system are fully described in Paper~I. Here we summarize some key features
of the data, which are useful to the present discussion.

The target GCs have been selected on the basis of two
requirements. First, a high metallicity in order to extend the model
calibration towards the range most relevant to ellipticals. Among the
12 clusters in the sample, 7 clusters have metallicities
$\feh\gapprox-0.5$ (on the scale of Zinn \& West, 1984), the most
metal-rich ones being NGC~6528 and NGC~6553. The remaining 5 clusters
are more metal-poor, and were included in the sample to check
consistency with previous studies (e.g. Trager et al.~1998).  Second,
the availability of independent estimates of element abundances, total
metallicities and ages, in order to allow for the empirical
calibration of the synthetic indices. As already mentioned, estimates
of the metallicity and [$\alpha$/Fe] ratios are available for the two
well-studied clusters NGC~6553 and NGC~6528, which ensures a
meaningful model calibration around solar metallicity. For the
remaining clusters estimates of the metallicity in the Zinn \& West
scale and in the Carretta \& Gratton~(1997) scale are available. This
allows us to use the clusters to calibrate the metallicity scale of
the models in a relative sense. Nevertheless, it would clearly be
useful to extend the detailed elemental abundance determinations to
all the clusters in the sample.

Spe\-cial ca\-re has been paid to sub\-tract the
fo\-re\-ground/back\-ground light from the cluster's light. Indeed,
the field and clusters stellar population components in the Bulge
appear to be virtually coeval, and to span similar metallicity ranges
(Ortolani et al.~1995; Zoccali et al.~2002). The very similar, though
not identical, spectral energy distribution of the bulge light could
therefore introduce spurious effects on the measurement of the cluster
indices, if not adequately subtracted.

The luminosities sampled in the GCs and the bulge fields by the slit
of the spectrograph have been carefully evaluated in order to assess
the dependence of the indices on stochastic effects. The number of
stars that are expected to be detected in the various evolutionary
phases is proportional to the total sampled luminosity and can be
easily evaluated (Renzini \& Buzzoni~1986, Maraston~1998,
Renzini~1998). Therefore, for every cluster it has been checked
whether the sampled luminosity is dominated by few, very bright stars,
like RGB-tip or E-AGB stars (Table~2, Paper~I). This is important for
metallic Lick indices like \mg, TiO, NaD, which are very strong in
these stars. The uncertainties on the indices associated with the
stocastic fluctuation in the number of stars which contribute the
light in the relevant wavelength ranges, are included in the error
budget (see Paper~1).

\section[]{A note on the metallicity calibration of SSP models}
\label{note}

Some of the Lick indices were designed as metallicity indicators for
unresolved stellar populations, therefore to calibrate a model Lick
index means to check whether the model gives the observed value of the
index for a SSP whose age and composition are independently known. In
practice, GCs offer the best proxy to a SSP. However, some ambiguities
make such calibration not so straightforward.

\subsection[]{Ambiguities in the definition of metallicity}
\label{note1}

From the model side, the total metallicity of model Lick indices is
not well defined because of the r\^ole of the so-called ``fitting
functions'' (see Sect. 5.1). The fitting functions are best fits of
the Lick indices as measured in stars, as functions of the stellar
parameters \teff, $g$~and chemical composition. According to the
standard procedure (e.g. Buzzoni et al.~1992, 1994; Worthey~1994), the
fitting functions are plugged on the isochrones to compute the Lick
indices of SSP models (Sect. 5, Eqs. 4-5). Therefore, it is
necessary to specify the metallicity parameter(s) for the fitting
functions and for the isochrones, and of course they should be the
same. While the latter is well-defined by construction (stellar
evolutionary models are constructed for well defined sets of
abundances), the former is somewhat ambiguous. Indeed, the estimates
of the chemical composition of the stars used for the fitting
functions come from a variety of sources, both spectroscopic and
photometric. These, quite inhomogeneous, metallicities are collected
under a parameter referred to as \feh~in the fitting functions
available in the literature (Worthey et al.~1994; Buzzoni et al.~1992;
1994). An additional source of complication comes from the fact that a
certain total metallicity might be achieved with different proportions
of the major elements, the so-called $\alpha$-elements, with respect
to iron. The fact that the fitting functions are derived from observed
stars implies that the element abundance ratios are not constant in
the fitting. In fact, as well known, the \afe~ratios vary
systematically among Milky Way stars (e.g. Mc William~1997), including
those in the samples used to construct the fitting functions
themselves. On the other hand the specific abundances of magnesium and
iron, beside the total metallicity likely affect the strength of Mg
and Fe absorption lines (see Sect.~4).

From the GC data side, the empirical metallicity scale of GCs is not
rigorously defined either. The reference values for the chemical
composition of the sample GCs used in this work are taken from the
revised compilation by Harris~(1996), which is largely based on the
Zinn \& West~(1984) scale. Thomas et al.~(2002b) show that the Zinn \&
West~(1984) scale, which is named as \feh, is likely to be closer to
the total metallicity rather than to the sole iron abundance. In fact,
the Zinn \& West scale is tied to the scale set up by Cohen~(1983), in
which the metallicities which are called \feh~are indeed obtained by
averaging [Mg/H] and [Fe/H] (Thomas et al.~2002b). This fact was
anticipated by the evidence that the integrated colours of SSP models
as function of the model total metallicity match well with those of
Milky Way GCs, when the metallicities of the latter are on the Zinn \&
West scale (Maraston~2000, Fig.~1). Moreover, the values reported in
the Harris~(1996) catalogue are {\it not} just the metallicity in the
Zinn \& West ~scale in all cases. When various estimates of
metallicity from other sources are available, either from spectroscopy
or from colour magnitude diagrams, these are used together with the
Zinn \& West-based value, and the straight average of the values is
published as \feh.

\subsection[]{The effect of the adopted Lick system}
\label{note2}

A quantitative model calibration requires that data and models refer
to the same spectro-photometric system (see Maraston et al.~2001a). The
definition of the Lick system (index passbands, resolution, etc.)  has
been slightly changed from Burstein et al.~(1984), through Worthey et
al.~(1994, hereafter W94) to Trager et al.~(1998). The different index
definitions may introduce offsets, which could affect the model
calibration. We adopt here the data as measured in the W94 version of
the Lick system, since the models are locked to this version via the
index fitting functions.  The effect of this choice has been tested by
computing the W94-like indices for the GC spectra obtained by Covino
et al.~(1995), then comparing them to the Covino et al. values of the
indices that are measured in the Burstein et al.~(1984) system. The
comparison for the case of 47~Tuc is shown in Table~1.
\begin{table}
\begin{minipage}{85mm} \caption{Lick indices of 47~Tuc
  (spectra of Covino et al.~1995) measured in the Worthey et al.~1994
  (W94) system for this work. The corresponding values in the Burstein et 
   al.~1984 (B84) system by Covino et al.~(1995) are given in
  second line.}  \begin{tabular}{cccccc}
\hline
\hline
\noalign{\smallskip}
Lick System & \mg  & \mgb & Fe5270 &  Fe5335  & \hb \\ 
 & (mag) & (\AA) & (\AA) & (\AA) & (\AA) \\ 
\hline
W94 & 0.15 & 2.19 & 1.64 & 1.40 & 1.48 \\ 
B84 & 0.18 & 3.02 & 2.18 & 1.88 & 1.62\\
\noalign{\smallskip}
\hline
\end{tabular}
\end{minipage}
\end{table}
The differences among the indices are not negligible. The values on
the W94 system are systematically lower than those on the Burstein et
al.~(84). In Sect.~\ref{calibration} we demonstrate the impact of
the index definitions on the model calibration.

\section[]{Results}
\label{results}
Fig.~1 shows the \mgb~ vs. the average iron
index~\fe~\footnote{\fe=(Fe5270+Fe5335)/2} of the GCs of our sample
(large filled symbols). The large open circle refers to the coaddition
of the spectra of 15 Bulge fields located in Baade's Window.  The blue
lines are models of Simple Stellar Populations (SSPs), i.e. coeval and
chemically homogeneous stellar populations, with total metallicities
(\zh)
\footnote{The notation \zh~is used to indicate total metallicities,
i.e. the total abundance of heavy elements with respect to hydrogen
normalized to the solar values,
i.e. $\zh=\log({{\mathrm{Z}}\over{\zsun}})-\log({{\mathrm{H}}\over
\hsun})$. By $\feh$~we mean the abundance of iron with respect to
hydrogen normalized to the solar values,
i.e. $\feh=\log({{\mathrm{Fe}}\over{\fesun}})-\log({{\mathrm{H}}\over
\hsun})$. If elements have solar proportions then $\feh=\zh$. In case
of $\alpha$-element enhancement, the relation between $\feh$~and
$\zh$~is: $\feh=\zh-0.94*(\afe)$~(Thomas et al.~2002b; see also Trager
et al.~2000).}
ranging from $-2.25$ to $+0.67$, and ages between 3 and 15 Gyr.  The
models are computed with the evolutionary population synthesis code of
Maraston~(1998), as described in Maraston \& Thomas~(2000; see also
Maraston et al.~2001b and Maraston~2002), and are based on
stellar tracks, implemented with the Worthey et al.~(1994) fitting
functions, to describe the stellar indices as functions of effective
temperature, gravity and metallicity. The stellar evolutionary tracks
are from Bono et al. (1997) and Cassisi et al. (1999) for
metallicities up to $Z=0.04$, from Salasnich et al. (2000) for
$Z=0.07$, and adopt solar abundance ratios. In the following, we refer
to these models as standard SSPs. \footnote{We want to emphasise that
what we call standard SSPs, i.e. those based on the Worthey et al. or
on the Buzzoni et al. fitting functions {\it are not} solar-scaled
SSPs at every metallicity. Indeed this type of models are constructed
by adopting the stellar indices of Milky Way stars, which have a
variety of abundance ratios, see Sect.~4.}

The GCs indices define a nice sequence to which the Bulge field
appears to belong as well. The sequence runs with a shallower slope
compared to the standard models, i.e. at a given \fe~index the data
have a stronger \mgb~than the models.  Several observational evidences
show that all galactic GCs have $\alpha$ enhanced abundance ratios,
with typical values around $\afe\sim+0.3$ (e.g. Pilachowski et
al.~1983; Gratton, 1987; Gratton \& Ortolani~1989, Carney 1996,
Salaris \& Cassisi 1996).  In particular, for the two most metal rich
clusters (NGC 6553 and NGC 6228 with $\mgb\sim3.8$) Barbuy et
al. (1999) find $\afe\sim+0.3$ from individual star spectroscopy.  The
Bulge field stars are also known to be overabundant in Mg with respect
to the solar ratio (McWilliam and Rich~1994).

Therefore the GC sequence traces the locus of $\alpha$-enhanced SSPs.

\begin{figure*}
\begin{minipage}{\textwidth}
 \begin{center}
 \psfig{figure=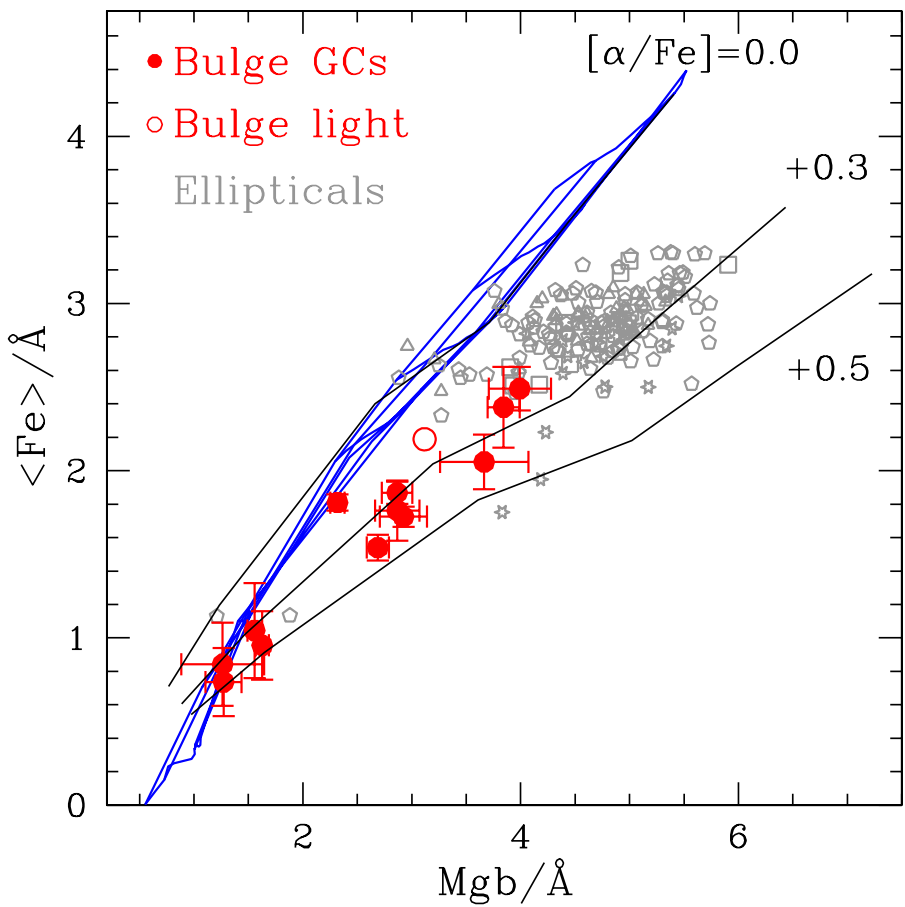,width=0.7\linewidth}
  \end{center}	
\end{minipage}
\caption{\mgb~indices vs. average iron \fe~indices of galactic and
Bulge GCs of our sample (large red filled symbols). The open circle
shows the average value of 15 Bulge fields located in Baade's
window. Small open grey symbols show the central values of the indices
for ellipticals taken from the literature: the field and cluster
ellipticals from Beuing et al.~(2002, pentagons); the Coma ellipticals
from Mehlert et al.~(2000, stars); the Fornax ellipticals of
Kuntschner \& Davies (1998, squares); the field and Virgo ellipticals
of Gonz\'alez~(1993, triangles).  Standard SSP models with
metallicities $\zh=(-2.25,~-1.35,~-0.55,~-0.33;~0.00;~+0.35;+0.67)$,
from bottom to top, and ages $3$,~$5$,~$10$,~$12$~and~$15$~Gyr, from
left to right, are shown as blue lines. The thick black lines show
12-Gyr SSP models with same total metallicities, and various
$\afe=0,~+0.3,~+0.5$~(Thomas et al.~2002b).}
\label{main}
\end{figure*}

Fig.~\ref{main} also shows the central values~(i.e. those obtained
with apertures $R\sim1/8\div1/10~R_{e}$) of the indices of field and
cluster ellipticals taken from various sources in the literature (see
caption).  The indices of ellipticals occupy a relatively narrow range
in \fe~and a large range in \mgb, stretching from the standard models
to the high metallicity extension of the GCs sequence, and
beyond. With very few exceptions, both \mgb~and \fe~indices are
measured stronger in the nuclei of ellipticals than in the most metal
rich GCs in our sample.  The stellar populations in the centers of
ellipticals seem to be characterized by:

(i) a supersolar total metallicity;

(ii) a range in abundance ratios, from almost solar, to \afe~values as
large as those of the most metal rich bulge clusters, or even more.

Similar conclusions have been proposed in the literature (Worthey et
al. 1992; references in Introduction).  Yet, these were based on the
assumptions that standard models reproduce the indices of solar
abundance ratios SSPs (at least at solar metallicity and above), that
\mgb~and \fe~trace the Mg and Fe abundance, and that an $\alpha$-
element overabundance affects the indices in the appropriate
direction.

Our comparison of the GCs data with the ellipticals, and especially the
inclusion of the indices of the metal rich clusters in the galactic
Bulge and the bulge field, confirm the validity of these assumptions,
the clusters being used as empirical SSPs.

This empirical evidence motivates the construction of new SSP models
with various, well-defined \afe~ratios which are shown in
Fig.~\ref{main} as thick black lines (Thomas et al.~2002b, hereafter
TMB02). The GCs are now very well represented by a coeval (12~Gyr old)
sequence of models with various metallicities and $\afe=+0.3$, in
agreement with the results from stellar spectroscopy. Note that at low
metallicities ($\mgb <2$) the standard models (blue lines) match with
the enhanced models. This is due to the standard calibrations by
Worthey et al.~(1994) being $\alpha$-enhanced at sub-solar
metallicities (Sect.~5; see also TMB02).

In the next Sect. we conduct a thorough analysis of the standard SSP
models, with the aim of assessing whether effects other than an
enhanced \afe~ratio can explain the deviation of the data from the
standard models. In other words, we investigate the uniqueness of the
``magnesium overabundance'' solution.


\section[]{Model Lick indices: key ingredients and ambiguities}
\label{modeling}
In the model SSPs, the line-strength of an absorption line with
bandpass $\Delta$ \issp~is given by
\begin{equation}
I=\Delta\cdot(1-{F{\rm_l}/F{\rm_c}}),
\end{equation}
where $F{\rm_l}$~and $F{\rm_c}$~are the fluxes in the absorption line
and in the continuum, respectively. 

In case of Lick indices this formula cannot be applied
straightforwardly because of the different spectral resolution of the
Lick system ($\sim 8$ \AA) and of the model atmospheres (Kurucz-based
stellar atmospheres have resolutions of 20~\AA~in the optical region
where the Lick indices are defined). Basically, the problematic
quantity is the flux in the absorption line $F{\rm_l}$~, because it
depends on the spectral resolution.  To overcome this problem, the
Lick group has measured the Lick indices on observed stellar spectra
having the required resolution (Burstein et al.~1984; Faber et
al.~1985). Assigning to each star of their sample the values for the
stellar parameters surface gravity ($g$), effective temperature
(\teff) and chemical composition, they have constructed polynomial
best-fitting functions which describe the various Lick indices
measured on the stars, \istarnoi, as a function of these parameters,
i.e. $\istarnoi={\mathrm {f}}(\teff; g; \feh)$. These polynomial
fittings following Gorgas et al.~(1993) are called fitting functions
(hereafter FFs).

The integrated Lick index of an SSP model is then evaluated as it
follows. 

The flux in the absorption line of the generic $\mathrm{i}$-th star of the
SSP, ${F{\rm_{l,i}^*}}$~, can be expressed with Eq.~1 as:
\begin{equation}
F{\rm_{l,i}^*}=F{\rm_{c,i}^*}\cdot(1-\istar/\Delta)
\end{equation}
where \istar~ that is the index of the $i$-th star is computed by
inserting in the FFs the values of (\teff;$g$; chemical composition)
of the $i$-th star, and \Fcstar~is its continuum flux. The latter is
computed by linearly interpolating to the central wavelength of the
absorption line, the fluxes at the midpoints of the red and blue
pseudocontinua flanked to the line (Table~1 in Worthey et al.~1994).
Eq.~1 can be re-written as:
\begin{equation} 
 \issp=\Delta\cdot(1-{{\sum_{\rm i}F{\rm_{l,i}^*}}/{\sum_{\rm i}F{\rm_{c,i}^*}}})=\bf {{\sum_{\rm i}\istari}}\cdot{f_{c,i}^{*}}
\end{equation}
where $f_{\mathrm{c,i}}^{*}$~are the contribution of each individual
star to the total continuum flux of the SSP.  Thus, the SSP integrated
index \issp~is the weighted average of the stellar indices
\istarnoi~with the weigths being $f_{\mathrm{c,i}}^{*}$.  When
computing actual models, the isochrone representing the SSP is binned
in \teff~subphases, small enough to ensure that \istari~is the same
for the stars belonging to the given subphase. A good binning is
$\Delta\teff\sim100~\mathrm {K}$~(Maraston~1998). Eq.~3 can
re-expressed for the subphases $j$
\begin{equation} 
 \issp={{\sum_{\rm j}\istarj}}\cdot{f_{c,j}^*}
\end{equation}
where $f^*_{\mathrm{c,j}}=\sum_{\mathrm i\in j} f_{\mathrm{c,i}}^{*}$.

It should be noted that since the r\^ole of the stellar continua is
that of a weight, it is not crucial that they are evaluated on
Kurucz-type spectra. A relation similar to Eq.~4 holds for
indices measured in magnitudes (e.g. \mg):
\begin{equation} 
 10^{-0.4\cdot\mg^{\rm SSP}}={{\sum_{\rm j}{\rm
10^{-0.4\cdot\mg^{*}{{\rm_j}}}}}} \cdot{f_{c,j}^*}
\end{equation}
The two ingredients (\istarj~and $f_{c,j}^*$) are discussed comprehensively
in the following subsections.
\subsection[]{Interplay between fitting functions \istarnoi and continua} 
\label{ffs}

To explore the systematic effects introduced in SSP models by the use
of different sets of FFs, and following Maraston et al.~(2001b), we
compute the same SSP models with three formulations for the FFs from
the literature, i.e.  by Worthey et al. (1994, hereafter Worthey et
al. FFs), Buzzoni et al. (1992, 1994, hereafter Buzzoni et al. FFs),
and Borges et al. (1995, hereafter Borges et al. FFs).

Worthey et al. FFs\footnote{We notice a typo in Table~3 of Worthey et
al.~1994. The fitting functions are mistakenly given as function of
$\log \theta$, while they are function of $\theta$, as also stated in
the text.}, the most widely used in the SSP models in the literature,
are based on the Lick sample of $\sim$~400 nearby stars.  Buzzoni et
al. FFs are based on a smaller sample of stars ($\sim$~87), also
located in the solar vicinity. As is well known, the $\alpha$ to Fe
abundance ratios in nearby stars vary with metallicity, ranging from
the super-solar values in the Halo stars ($\afe\sim+0.3$) to the solar
proportions in the metal-rich disk stars ($\afe=0.0$) (e.g. Wheeler et
al.~1989; Edwardsson et al.~1993; Fuhrmann et al.~1995; Fuhrmann,
1998; see the comprehensive review by McWilliam~1997). Thus, likely,
these two sets of FFs reflect $\alpha$ enhanced mixtures at low $Z$,
and solar ratios at high metallicity, a trend which is dragged into
the SSP models through \istarj ~(Eqs.~4-5).  This explains why the
standard models in Fig.~\ref{main} (blue lines) represent well the
indices of metal poor GCs.

Borges et al. FFs (see also Idiart \& de Freitas Pacheco~1995) have
been derived from a sample of roughly 90 stars for which the Mg to Fe
ratio has been measured. Thus, they include the [Mg/Fe] ratio as an
additional variable besides temperature, gravity and metallicity.

In the following subsections we discuss separately high metallicity
(Sect. 5.1.1) and low metallicity (Sect. 5.1.2) model indices.

\subsubsection[]{{\bf The metal-rich zone}} 
\label{ffshighz}
\begin{figure*}[!ht] 
 \psfig{figure=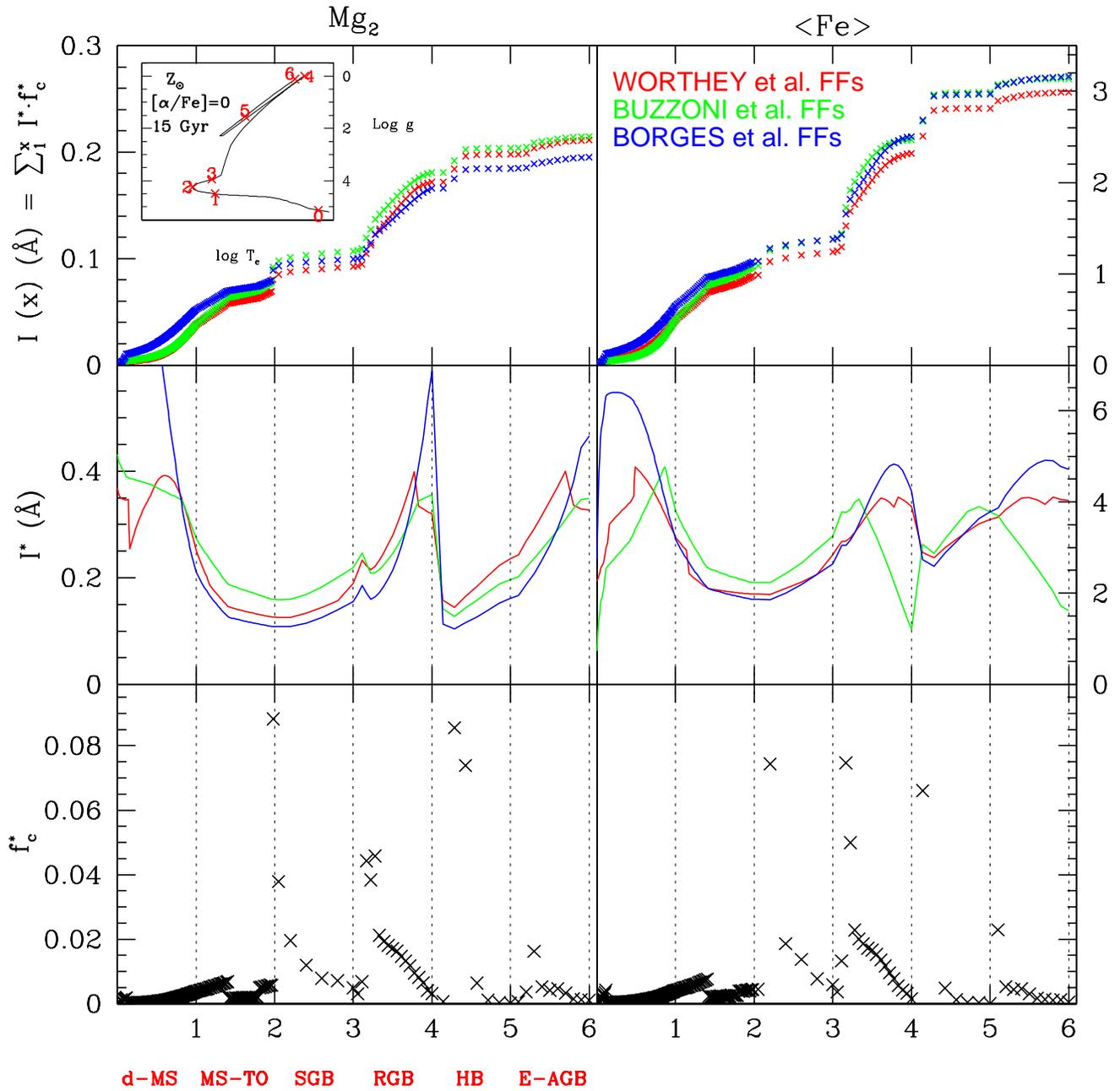,width=\linewidth}
\caption{Closer look to model \mg~(left-hand panels) and
\fe~(right-hand panels) indices of a 15 Gyr old SSP with solar
metallicity, solar abundance ratios, and Salpeter IMF.  The upper
panels show the cumulative SSP indices $I^{\mathrm{SSP}}(x)$, integrated up
to the point $x$ along the isochrone (see text): $x=1$: lower Main
Sequence (d-MS) up to $\teff \leq 5000~{\mathrm{K}}$; $x=2$: Main Sequence
up to Turn Off (TO); $x=3$: Sub Giant Branch (SGB); $x=4$: Red Giant
Branch (RGB); $x=5$: Horizontal Branch (HB); $x=6$: Early Asymptotic
Giant Branch (E-AGB). These phases are identified on the corresponding
isochrone (small inserted panel), shown in the temperature-gravity
plane. Here the label $\afe=0$~refers to the composition of the
stellar tracks.  Colors code the different FFs for \istarj~(Eq.~4-5)
adopted for the SSP models, by Worthey et al. (red lines), Buzzoni et
al. (green lines) and Borges et al. (blue lines). The values
\istarj~are shown in the middle panels, with lines connecting the
stellar indices of the subphases along the isochrone. Finally the
lower panels display the contributions of the various subphases to the
total continuum flux, \fcstarj~(Eq.~4-5). Note that the flux
contributions in post-MS ($x>2$) take into account the fuel
consumption, i.e. the product of the stellar lifetimes with the
stellar luminosities. For example the contribution of the long lasting
RGB bump phase (peak in phase 3-4) is much larger than that of the
RGB-tip (end of phase 4), where, in spite of larger stellar
luminosities, the evolutionary timescale is much shorter. }
\label{buildupzsun}
\end{figure*}

Fig.~\ref{buildupzsun} il\-lu\-stra\-tes the in\-ter\-play be\-tween
\istarj~and \fcstarj~(E\-qua\-tion~4-5) in de\-ter\-mi\-ning the SSP
magnesium and iron indices in the particular case of a 15 Gyr, solar
metallicity and solar abundance ratio SSP, with Salpeter IMF.  The
three sets of FFs considered are color coded as marked in the
top-right panel.  The $x$-axis is a monothonic coordinate along the
SSP isochrone. The integer values of $x$~(1 to 6) mark the end of the
six main evolutionary phases (see the caption and the figure with the
isochrone inserted in the top-left panel).  Each $x$-point in Main
Sequence ($x\leq2$) represents the subphase along the Main Sequence
isochrone. Each $x$-point in post-MS ($x>2$) represents the $j$
subphases of every post-MS major phase, which are equally spaced in
effective temperature ($\Delta\teff\sim 100~{\mathrm{K}}$~,
Maraston~1998).
The top panels show the cumulative \mg($x$)\footnote{Notice that we
plot the \mg~index expressed in \AA,
i.e. $\textrm{Mg}_2(\textrm{\AA})=1-10^{(-0.4\cdot\textrm{Mg}_2(\textrm{mag}))}$.
We use \mg~ here instead of \mgb~ because neither Buzzoni, nor Borges
give FFs for \mgb. The two indices, however, are very closely
related.}  (left) and \fe($x$) (right) along the isochrone, which
assumes the value of the SSP model at $x=6$. As in Eq.~4-5, each value
of \mg($x$) and \fe($x$) is obtained by summing up to point $x$, the
product of \istarj~times \fcstarj. These are separately shown in the
central and lower panels, respectively. The behavior of \istarj~in the
central panels reflects the changing of \teff~along the isochrone,
both \mg~and\fe~ indices being very strong in cool stars.

The three sets of FFs correspond to quite different values for
\istarnoi~along the isochrone, particularly in the faint dwarf regime
($x\lapprox 1$), at the Tip of the RGB ($x \sim 4$), and at the end of
the E-AGB ($x \sim 6$).  In spite of that, the SSP indices keep very
close along the isochrone, and assume quite similar total values, due
to the low contribution to the total continuum flux of these
particular phases.  Thus, the indices turn out to be quite insensitive
to the adopted set of FFs.

As a result of the weighting through \fcstarj, the lower MS, RGB and
E-AGB bright stars (which have very strong \mg) are not important in
determining the total SSP index.  The most important contributors to
the continuum fluxes in the two windows of \mg~and \fe~are: the stars
around the TO; those on the fainter portion of the RGB (especially at
the so-called bump), and the Horizontal Branch stars.  Their indices
dominate the integrated values.  This applies in general to old
($\gapprox~3~{\mathrm{Gyr}}$) stellar populations, as the relative
contribution to the total optical flux of the different phases does
not depend much on age in this age range (Renzini \& Buzzoni~1986,
Maraston 1998).  
\begin{table}
 \centering \begin{minipage}{65mm} \caption{Relative contributions of
  stellar evolutionary phases, to the continuum flux of the \mg~index
  ($\lambda \sim 5175$\AA) and \fe~($\lambda \sim 5300$\AA), for
  15-Gyr old SSPs with Salpeter IMF, and metallicities: \zsun~and
  $Z=10^{-4}$. RGB tip is the portion within 1 mag from the tip.}
  \begin{tabular}{lccccc} 
  & \multicolumn{2}{c}{\zsun} & \multicolumn{2}{c}{$10^{-4}$\zsun} \\
\noalign{\smallskip} \hline \hline \noalign{\smallskip} phase & \mg &
\fe & \mg & \fe \\ \noalign{\smallskip} \hline \hline
\noalign{\smallskip} d-MS & 0.11 & 0.13 & 0.10 & 0.10 \\ MS TO & 0.32
& 0.22 & 0.26 & 0.26 \\ SGB & 0.05 & 0.12 & 0.10 & 0.10 \\ RGB & 0.29
& 0.30 & 0.28 & 0.28 \\ RGB-tip & 0.01 & 0.01 & 0.10 & 0.10 \\ HB &
0.17 & 0.17 & 0.12 & 0.12 \\ E-AGB & 0.05 & 0.05 & 0.04 & 0.04 \\
\noalign{\smallskip} \hline
\end{tabular}
\end{minipage}
\end{table}

The contributions to the total continua of \mg~and\fe~of the various
evolutionary phases are given in Table~2 for 15 Gyr old metal-rich and
metal-poor SSPs.

As apparent in Fig.~\ref{buildupzsun}, Borges et al. FFs provide
\istarj~for faint dwarfs which are much larger than those of the other
two sets. These result from the exponential increase with decreasing
\teff~of their FF for \mg. In the validity range as specified by the
authors (\teff$\gapprox3800~{\mathrm{K}}$), the FF for \mg~yields values as
high as 1.9 mag (corresponding to 0.83 \AA~in the units of
Fig.~\ref{buildupzsun}), while the coolest dwarf in their observed
sample has $\mg=0.45$~mag (or 0.34 \AA). These very strong
(extrapolated) \mg~indices as obtained with a blind use of the FFs are
extremely unrealistic. This example illustrates the importance of
checking the behaviour of the algebraic FFs when computing SSP models.

We conclude that the Mg-Fe relation of standard models around solar
metallicity (Fig.~\ref{main}) is independent of the fitting
functions.


\subsubsection[]{{\bf The metal-poor zone}}
\label{metalpoor}

As shown in the previous paragraph, the differences in the FFs appear
to be unimportant at $\sim\zsun$. Actually, the \mg~index obtained
with the Borges et al. FFs does deviate from the other two values, by
an amount which is comparable to the typical observational error
affecting the GC data ($\sim 0.01$). This discrepancy becomes more
pronounced at very low metallicities, which is relevant for the model
calibration with GCs.
\begin{figure} 
 \psfig{figure=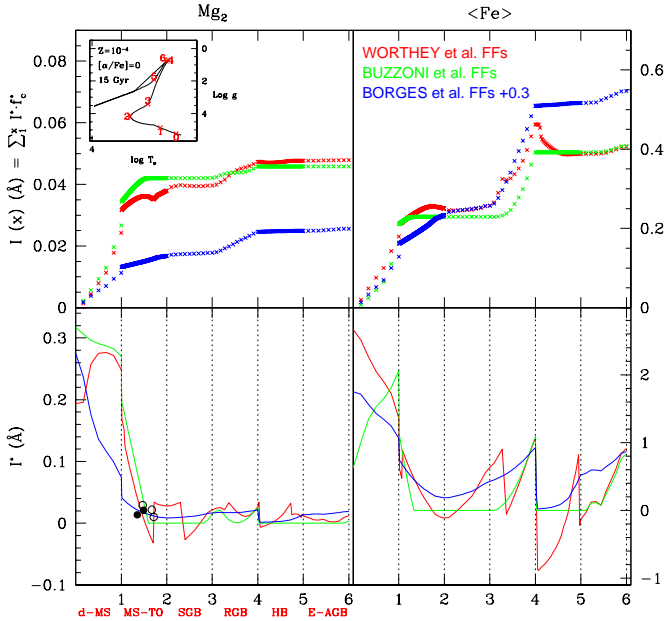,width=\linewidth}
\caption{Same as Fig.~2 but for a very metal-poor SSP with
metallicity $Z=10^{-4}$. Note that the integrated \mg~is made up by
the lower Main Sequence (d-MS, phase~1). The symbols in the left-hand
bottom panel are the values of \mg~of the stars in the Lick sample
(filled) and in the sample used by Borges (open) with metallicities
and gravities appropriate to the MS of these SSP models. As in
Fig.~\ref{buildupzsun}, the label $\afe=0$~in the inserted panel
refers to the composition of the stellar tracks.}
\label{builduplowz}
\end{figure}

Fig.~\ref{builduplowz} is the analogous of Fig.~\ref{buildupzsun},
but for a very metal-poor SSP with metallicity $Z=10^{-4}$.  We use
here the Borges et al. FFs with [Mg/Fe]=0.3 which is appropriate at
low metallicities.

At low me\-tal\-li\-ci\-ty the tem\-pe\-ra\-tu\-re di\-stri\-bu\-tion
of the i\-so\-chro\-ne shifts to hotter values (see inserted panels in
Fig.s~\ref{buildupzsun} and \ref{builduplowz}).  The relative
contributions to the continuum flux of the different evolutionary
phases is very similar to the \zsun~ case (see Table 2), but the
stellar indices are now very weak, and the strongest \mg~are found on
the lower MS, which is the coolest portion of the isochrone. As a
result, the total \mg~of the SSP is very close to the value attained
already at the turn-off point.  Borges et al. FFs yield much lower
stellar indices for the lower MS, compared to the other two FFs, which
explains the lower integrated index.  Incidentally, this is true also
when adopting [Mg/Fe]=0. in the FF formula.  It should be noted that
the behaviour of the FFs at the low Main Sequence (phase 1) is not
constrained by stellar data. Indeed, only 5 main sequence stars
(symbols in the lower left panel of Fig.~\ref{builduplowz}) are
found at $\feh\approx-1.8$ by merging both the Lick and the Borges et
al.~data base. An improvement of present low-metallicity SSP models
can be gained by implementing the stellar libraries with cool, dwarf
and low metallicity objects.

The importance of the (lower) main sequence phase on the integrated
\mg~has the consequence of making this index sensitive to the mass
function of the stellar population. This complicates the comparison of
the models with the GC data, for which the present day mass function
derives from the IMF plus the possible dynamical evolution, which can
lead to the evaporation of the low mass stars (e.g. Piotto \& Zoccali
1999).

Different from the \mg~index, an important contribution to the total
\fe~index comes from the RGB portion of the isochrone, where the FFs
are very noisy (see right panels in ~\ref{builduplowz}).  Notice that
the index computed with Worthey et al. FFs converges to that based on Buzzoni
FFs because of negative stellar indices predicted for the hot HB
stars.  It is very difficult to assess the reliability of the indices
as metallicity indicators at such low-$Z$. 

\subsection[]{IMF effects}

As discussed in the previous sections, \mg~indices are very strong in
dwarf stars, therefore dwarf-dominated stellar population could in
principle allow to reach the very high values of \mg~shown by
ellipticals.
\begin{figure} 
 \psfig{figure=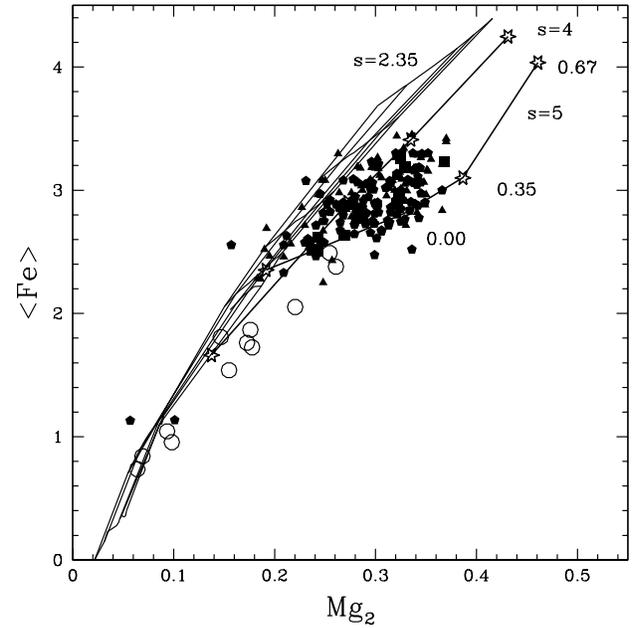,width=1.\linewidth}
\caption{Effect of a dwarf-dominated IMF on the \mg~and\fe~diagram of
SSPs.  The model grid for the Salpeter IMF exponent of 2.35 is the
same as in Fig.~\ref{main}.  The two additional solid lines are
dwarf-dominated SSPs with IMF's exponents of 4 and 5,
respectively. These SSPs have 12 Gyr and metallicities as indicated by
the labels. Data of GCs and ellipticals as in Fig.~\ref{main}.}
\label{imf}
\end{figure}
Fig.~\ref{imf} shows the location of such dwarf-dominated SSPs in
the \mg~vs.~\fe~diagram (lower solid lines). Lines connect 12 Gyr old
models with metallicities from 3 time to half solar, and IMF's
exponents of 4 and 5 (in the notation in which Salpeter is 2.35), from
top to bottom, respectively. Worthey et al. FFs have been used for
this exercise.  Data of ellipticals and GCs are the same as in
Fig.~\ref{main}.

Dwarf-dominated stellar populations are able to reproduce the \mg~and
\fe~indices of ellipticals without invoking abundance effects. It
should be noted that in such dwarf-dominated SSPs, $> 70\%$ of the
total luminosity is made up by stars close to the H-burning limit.
The different slope of these SSP models on the \mg $-$ \fe~plane
reflects the different dependence of the stellar indices from \teff~at
the lower end of the MS.

These extreme IMFs are also able to reproduce the low values of the
Calcium triplet absorption line at 8600~K observed in ellipticals
(Saglia et al.~2002), because its strength decreases with increasing
gravity (Jones et al.~1984).  However, these models fail at
explaining other spectral properties of ellipticals. In fact the
corresponding stellar mass-to-light ratios become much larger
($M/L_{B}>~30$, see Maraston~1998) than the dynamical ones observed in
the central portions of ellipticals ($M/L_{B}\sim6$, Gerhard et
al.~2001). Similarly, a dwarf-dominated elliptical galaxy light was
excluded given the strength of the CO absorption (Frogel et al.~1978)
and the absence of the Wing Ford bands in absorption (Whitford ~1977).
The case of such a dwarf-dominated present mass function for GCs, and
the Bulge field is ruled out by direct observations of the lower MS
(de Marchi \& Paresce~1997; Piotto \& Zoccali~1999; Zoccali et
al.~2000a).

We conclude that an extremely steep IMF is not a viable alternative to
explain the high values of the Mg indices in ellipticals.

\subsection[]{The effect of stellar evolutionary tracks}

As already stated, our standard SSP models are based on the stellar
tracks by Cassisi et al.~(1999). Since differences exist among
different sets of tracks, it is interesting to check their impact on
the model indices. To this aim we have computed SSP models with
Worthey et al. FFs, but varying the input tracks. The Padova stellar
tracks and isochrones as available on the Web have been used,
specifically those by Fagotto et al.~(1994) for metallicities
$\zh=-1.69$ and $-0.69$, and those by Salasnich et al.~(2000), with
solar scaled abundance ratios, for metallicities $\zh=-0.4, 0, 0.35$.
The fuel consumption theorem is adopted also for these SSPs, following
the method described in Maraston~(1998;~2003).
\begin{figure} 
 \psfig{figure=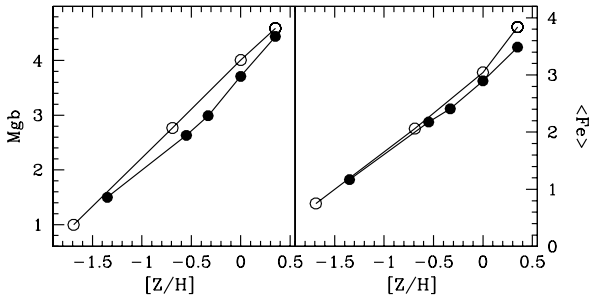,width=\linewidth}
\caption{Effect of stellar tracks on the Lick indices of 12 Gyr SSPs
models with various total metallicities \zh. Filled symbols: SSPs
adopting the Cassisi tracks, open symbols: SSPs adopting the Padova
tracks.}
\label{tracks}
\end{figure}
The results are shown in Fig.~\ref{tracks}, which compares the
\mgb~and~\fe~indices of 12 Gyr SSP models as a function of the total
metallicities \zh, as obtained with the Cassisi tracks (filled
circles) and the Padova tracks (open circles),

The use of the Padova tracks produces slightly stronger \mgb~indices,
at metallicities \zh~between $\sim -0.5$ and solar. This effect is
most likely due to the cooler temperatures of the Padova tracks along
the Red Giant Branch with respect to the Cassisi tracks (see also
Maraston~2003). A very modest impact is present also for the
\fe~index.  Note that at metallicities above solar the \fe~obtained
with the Padova tracks is stronger than that obtained with the Cassisi
tracks, while the corresponding \mgb~indices are consistent. This
implies that the discrepancy with ellipticals data is slightly larger
when the Padova tracks are used. The differences are however very
small, and the conclusions drawn from Fig.~\ref{main} are not
affected by our use of a specific set of stellar tracks.

Re\-cen\-tly, i\-so\-chro\-nes and tra\-cks with su\-per-so\-lar
\afe~ra\-tios be\-ca\-me a\-vai\-la\-ble (Sa\-las\-nich et al.~2000
wi\-th $\afe=+0.3$; see al\-so Berg\-busch \& Van\-den\-Berg 2001; Kim
\- et \- al. 2002). The\-se im\-pro\-ved u\-pon pre\-vious
cal\-cu\-la\-tions that con\-si\-de\-red only the effect on nu\-clear
rea\-ction ra\-tes (e.g.~Sa\-la\-ris et al.~1993), whi\-le the
up\-da\-ted tra\-cks ha\-ve al\-so in\-clu\-ded the ef\-fect of
$\alpha$-en\-han\-ce\-ment on the stel\-lar o\-pa\-ci\-ties. He\-re we
use the Sa\-las\-nich's com\-pu\-ta\-tions in or\-der to che\-ck the
im\-pact of $\alpha$-en\-han\-ced tra\-cks on the fi\-nal in\-dex
va\-lues of SSP mo\-de\-ls. For con\-si\-sten\-cy, we com\-pa\-re the
in\-di\-ces ba\-sed on the two Pa\-do\-va sets, with so\-lar sca\-led
and $\alpha$-en\-han\-ced a\-bun\-dan\-ce ra\-tio.
\begin{figure} 
 \psfig{figure=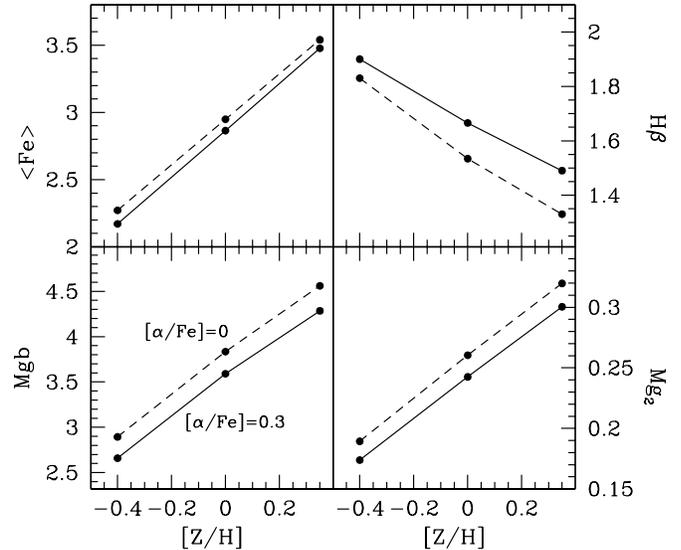,width=\linewidth}
\caption{Effect of $\alpha$-enhanced stellar tracks on \mg~, \mgb,
\fe~and \hb~of 10 Gyr SSPs (Salpeter IMF) with various total
metallicities \zh. Dotted lines: solar-scaled tracks. Solid lines:
$\afe=+0.3$~tracks. Tracks are from Salasnich et al.~(2000), the
Worthey FFS are adopted.}
\label{sala}
\end{figure}
This is shown in Fig.~\ref{sala} for the illustrative case of 10 Gyr
old SSPs\footnote{We use here 10 Gyr to avoid interpolation on the Salasnich
et al. isochrones}.

For the same total metallicity \zh, the Mg and Fe indices of SSPs
based on $\alpha$-enhanced stellar tracks (solid lines) are {\it
lower} than those based on solar scaled stellar tracks (dotted lines),
This happens because the $\alpha$-enhanced tracks are bluer than the
solar-scaled ones at fixed total metallicity (Salasnich et al.~2000),
because of the lower stellar opacities. So, by increasing \afe~ {\it
at constant} \zh~the Mg indices actually decrease!

This rather counterintuitive behavior is a consequence of the fact
that, at given total metallicity, the increase of the \afe~ratio
produces only a small increase of the $\alpha$ elements, and instead a
large decrease of iron. In fact, in solar proportions the total
metallicity is by far dominated by the $\alpha$-elements
($\sim74\%$~by mass), the iron-peak elements amounting only to
$\sim8\%$, the residual being contributed by elements produced by the
p, s, and r processes (Trager et al.~2000; TMB02).For
example, compared to solar proportion in a mixture with \afe=+0.3 the
$\alpha$ elements will contribute slightly more than $\sim74\%$, but
the iron peak elements will be reduced by almost a factor of two, down
to $\sim4\%$. Therefore, the main effect is to decrease iron rather
than to increase, e.g., magnesium. Since iron is the most effective
electron donor (e.g. Salasnich et al.~2000), the lower abundance of
iron in enhanced \afe~mixtures has the effect of decreasing their
low-temperature opacities, which in turn determine an increase of the
temperature of the RGB, and finally a decrease of the Mg indices
because these are stronger in cool stars.

The {\it bluing} of the isochrone also affects the \hb~line indices,
which are {\it stronger} when $\alpha$-enhanced tracks are adopted
(Fig.~\ref{sala}, upper right panel). Therefore, in order to
reproduce the observed \mg, \fe~ and \hb~ indices, the
$\alpha$-enhanced tracks require {\it older} ages.

It should be noted that for the models of Fig.~\ref{sala}, the
parameter expressing the chemical composition in the Worthey et
al. FFs (referred to as \feh~in Worthey et al.~1994) has been
considered to represent the total metallicity. As discussed in
Sect.~\ref{note}, this might be not entirely correct, since a
variety of methods have been used, both photometric and spectroscopic,
to determine such parameter for the stars of the Lick sample. If the
\feh's values are used, the metallic indices shown in
Fig.~\ref{sala} decrease even further.

Fig.~\ref{sala} illustrates that accounting {\it only} for
$\alpha$-enhanced stellar tracks to compute $\alpha$-enhanced SSP
models, while using the same FFs, affects the indices in the wrong
direction with respect to the locus occupied by the metal-rich GCs and
ellipticals data in Fig.~\ref{main} (\mg~is affected more than \fe).
A fully consistent exploration of the effect of the \afe~ratio
requires also the use of fitting functions depending on \afe. This is
done in TMB02.

%

\subsection[]{Summary}
\label{summary}

The conclusions of the previous paragraphs are the following.

1. Differences in the available sets of FFs do not affect the integrated
\mg~ and \fe~indices, due to the low contribution to the total continuum
flux of those evolutionary phases where the FFs are mostly discrepant.

2. For both \mg~and \fe~(and in general indices measured in the
optical) the most important contributors are MS TO, RGB and HB
stars. These are the evolutionary phases where the FFs need to be best
constrained from stellar data.

3. Results 1 and 2 hold for metallicities $\feh \gapprox -1$, and are
largely independent of the age and stellar tracks used.

4. At very low metallicities (less than \zsun/10) the lower MS appears
to dominate the value of the SSP \mg~index. Uncertainties in the FFs
and in the mass function jeopardize the calibration of the theoretical
indices with the GC data.
 
5. At metallicities \gapprox \zsun/2, the slope of the {\it solar
scaled} \mg~vs \fe~relation for SSP models seems quite robust. One
possibility to get strong \mg~indices in combination with weak \fe~
without invoking a super-solar \afe, is to adopt a very steep exponent
for the IMF. However, IMF exponents as large as 4 $\div$ 5 have to be
used in order to encompass the locus occupied by ellipticals. Such values are
ruled out by other constraints.

6. When $\alpha$ enhanced tracks are used the \mg~and \fe~indices become
$weaker$, due to the $blueing$ of the isochrone. Therefore, in order to
match the observational data of GCs and ellipticals, abundance effects have to be
accounted also in the FFs.

\section[]{Model Lick indices: comparison with the data}
\label{calibration}

In this section we compare quantitatively the indices for our sample
of GCs with the models. Good models have to fulfill two requirements:
i) the metallicities obtained using different line-strengths have to
be consistent; ii) the metallicities derived from the models have to
be in agreement with those determined independently from,
e.g. spectroscopy of stars in GCs or CMD fitting.  We already know
from Fig.~\ref{main} that condition i) will not be fulfilled at $\zh
\gapprox -0.6$, because the GCs data deviate from the models. It is
however important to check quantitatively the discrepancy.  For a
comparison of the standard SSP models used here with Magellanic Clouds
clusters, we refer to Beasley et al.~(2002).

In this section we also explore the significance of other Lick indices
as metallicity indicators. Finally we calibrate the Balmer lines.

\subsection[]{Chemical compositions from Mg and Fe indices}
\label{calibramg}

\begin{figure*}[!ht] 
\centering \includegraphics[width=\textwidth]{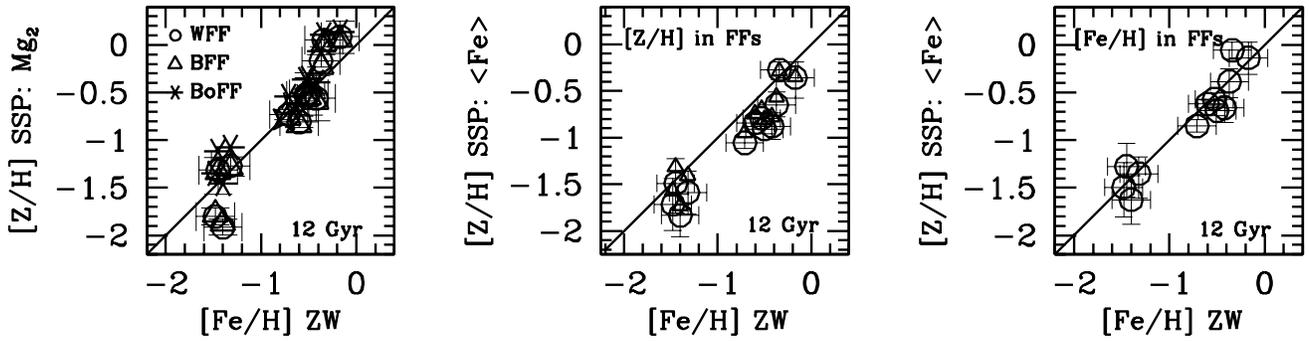}
\caption{Comparison between the SSP-derived total metallicities from
\mg~(left-hand panel) and \fe~(central panel) for our sample of GCs,
with the empirical metallicity scale \feh~as compiled by
Harris~(1996), which is largely based on the Zinn \& West~(1984)
scale. The standard SSPs, i.e. those based on Milky Way calibrated
FFs, used here adopt the Cassisi tracks and: the Worthey et al. FFs
(open circles), the Buzzoni et al. FFs (open triangles), the Borges et
al. FFs (asterisks). The age of these SSPs is 12~Gyr. The SSP-derived
\zh~is obtained by interpolating the observed indices on the model
grid, separately.  Diagonal lines show the 1 to 1 relations. In the
right hand-panel the \fe~index of the SSP models is derived by
plugging \feh~in the FFs instead of the total metallicity \zh (see
Sect.~3), for the only case of the Worthey et al. FFs.}
\label{calibra}
\end{figure*}
Fig.~\ref{calibra} compares the metallicities \zh~derived from the
standard SSP models, with those provided by the revised compilation of
Harris~(1996), which is largely based on the Zinn \& West~(1984)
scale, for each GC of our sample. The model \zh~is obtained by
interpolating the \mg~index (left-hand panel) and the \fe~index
(central panel) on the SSP models (12 Gyr) based on the Cassisi tracks
plus the FFs by: Worthey et al. (open circles), Buzzoni et
al. (triangles), and Borges et al. (asterisks). The errorbars connect
the minimum and maximum model metallicities obtained by subtracting
(adding) the observational errors to the measured values. For the
empirical metallicities, a conservative error of +0.2 dex has been
considered. The right-hand panel refers to a model \fe~index as
obtained by using as a metallicity input in the Worthey et al.~FFs,
the value of the iron abundance \feh~and assuming an $\afe=+0.3$ (see
Sect.~\ref{note}). Also in this case, the interpolation is with the
model total metallicity \zh.

Fig.~\ref{calibra} shows that for standard SSP models:

i) the total metallicities \zh~as derived from the \mg~index
(left-hand panel) are well consistent with the empirical scale of Zinn
\& West at metallicities lower than $\sim-0.5$. For NGC~6553 and NGC~6528
the models give \zh~values somewhat in excess of the values on the
Zinn \& West scale, but would agree with the near solar abundance
indicated by the spectroscopic observations (Barbuy et al.~1999; Cohen
et al.~1999).

ii) the total metallicities \zh~as derived from the \fe~index, when
the latter is computed with \zh~as input of the FFs (central panel),
are systematically lower than the empirical ones, by roughly 0.3~dex.

iii) the total metallicities \zh~as derived from the \fe~index, when
the latter is computed by using instead the iron abundance \feh~and
assuming a 0.3 dex $\alpha$-enhancement as input of the FFs
(right-hand panel) is consistent with the Zinn \& West values.

On the basis of these evidences we conclude that: 

i) the standard SSP models, i.e. those based on the Milky Way
calibrated FFs, do not underestimate the \mg~index; rather they
overestimate the \fe~index. The disagreement between the models and
the GC (and ellipticals) data would then point towards an {\it iron
deficiency}, as opposed to magnesium enhancement, at virtually all
metallicities. Suggestions in this direction can be found in Buzzoni
et al.~(1994), Trager et al.~(2000), TMB02;

ii) the model-derived {\it total} metallicities \zh~are in agreement
with the metallicities on the Zinn \& West scale which are referred as
to \feh~(see Sect.~3.1).

In the left-hand panel of Fig. \ref{calibra} two of the four
clusters at $\feh_{\mathrm{ZW}}\sim -1.4$~(NGC 6218 and NGC 6981) have a
too low \mg-derived metallicity, when the Worthey et al. or the
Buzzoni et al. FFs are used. The SSPs with the Borges et al. FFs,
instead reproduce the \mg~indices of these two specific clusters,
because of the lower \mg~indices of these FFs (see Sect. 4.2). Since
at low $Zs$ the \mg~index is dominated by the lower MS component
(Sect.~5.3), dynamical effects stripping low mass stars could be
responsible for the observed low \mg~indices of these particular
objects. Piotto et al.~(2001) show that indeed the mass function of
NGC~6981 is consistent with a power-law with a slope flatter than the
Salpeter one.  We note that NGC~6218 is the object of our sample with
the poorest sampled light (paper I).

\begin{figure*}[!ht] 
\centering \includegraphics[width=\textwidth]{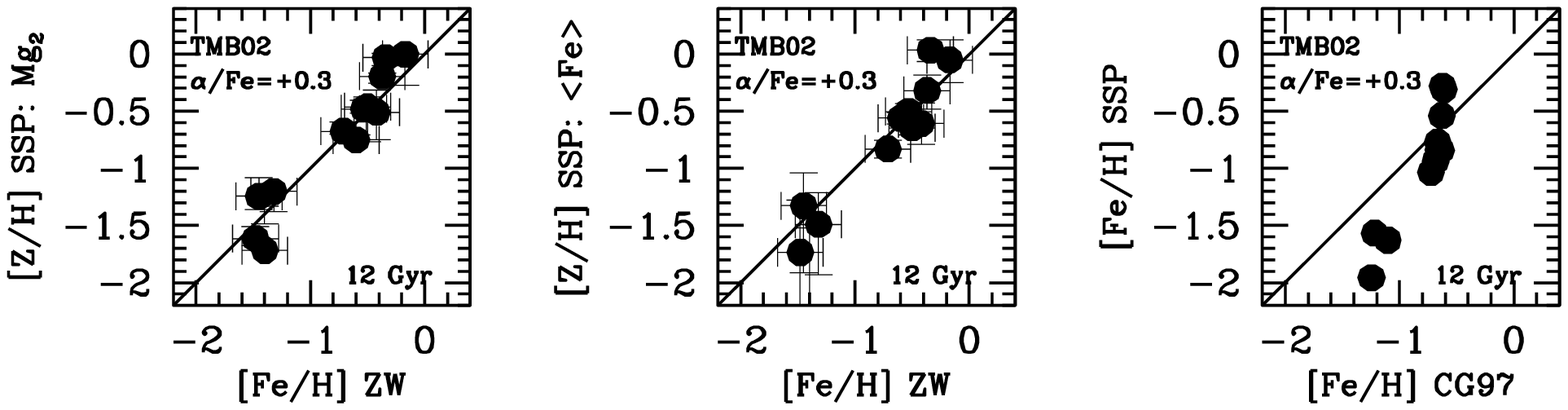}
\caption{The metallicities of the GCs as derived from the
$\alpha$-enhanced SSPs of TMB02, with age of 12 Gyr and $\afe=+0.3$. In
the left-hand and central panels, the total metallicity \zh~as derived
with the \mg~and the \fe~index, respectively are compared with the
values on the Zinn \& West scale. In the right-hand panel the
predicted \feh~is compared to the empirical scale of iron abundances
by Carretta \& Gratton~(1997).}
\label{tmb02}
\end{figure*}

Fig.~\ref{tmb02} compares the metallicities of each GCs as obtained
from the \mg~(left-hand panel) and \fe~(right-hand panel) index using
the TMB02~$\alpha$-enhanced SSP models, with 12 Gyr and
$\afe=+0.3$~(solid black lines in Fig.~\ref{main}). These models
include the dependence of the fitting functions for \istarnoi~on the
\afe~parameter.

The total metallicities \zh~as derived from \mg~and \fe~are in
excellent agreement with each other. This results from having taken a
super solar \afe~abundance ratio in the models into account. The
metallicities are in excellent agreement with those in the Zinn \&
West scale, over the whole range covered by our sample.  The assumed
$\afe=+0.3$ for the GCs is in agreement with the \afe~measured
spectroscopically (see e.g. Carney~1996; Salaris \& Cassisi~1996). For
the Bulge GC NGC~6528 recent spectroscopic abundance determinations
give: $Z\sim\zsun$, $\afe\sim0.3$, $\feh\sim-0.3$ (Barbuy et al.~1999;
Origlia et al.~2001). We acknowledge that some controversy exists in
the literature: Carretta et al.(2001; see also Cohen et al.~1999) give
$\feh\sim+0.08$ for 6528 and $\feh\sim-0.06$ for NGC~6553,
$\afe\sim0.2\div0.4$. It would be extremely important to pin down the
element abundances for these objects which are the most metal-rich
calibrators available.

Carretta \& Gratton~(1997) provide a metallicity scale for GCs which
should reflect the \feh~abundance. The right-hand panel of
Fig.~\ref{tmb02} compares the model-derived \feh~with this
scale. The metallicities in the Zinn \& West scale are transformed
into the Carretta \& Gratton scale by adopting the relation provided
by the authors. This comparison is rather poor. On this scale the iron
abundance of our GC sample appears clustered on two values, while the
observed \fe~indices span a considerable range. It seems difficult to
reconcile our index-based \feh~ with those in the Carretta \& Gratton
scale.

We conclude this section discussing the case of 47~Tuc, which allows
us to demonstrate the importance of comparing data and models which
are set on the same system (see Sect. 3.1, Table~1). If we use the
indices measured in the Burstein et al.~(1984) system, 47~Tuc has
$\zh\simeq-0.50$; if we use the indices measured in the Worthey et
al. one, we get $\zh\simeq-0.78$. This last value is in perfect
agreement with the metallicity of 47~Tuc in the Zinn \& West~scale
($\zh\sim-0.76$, from Harris~1996).  In addition we notice that the
metallicity derived for 47~Tuc using the indices in the Burstein et
al.~(1984) is as large as that of NGC~6356, while the comparison of
the RGB ridge lines of these two clusters indicates that 47 ~Tuc is
more metal-poor (Bica et al.~1994). For a critical discussion focused
on 47 Tuc see Schiavon et al.~(2002).
\begin{table*}[!ht]
\caption{The metallicities \zh~derived with \mg~and \fe~of standard
SSPs adopting the Worthey et al. FFs (col. 3-4), the Buzzoni et
al. FFs (col. 5-6) and the Borges et al. FFs (see left-hand and
central panel of Fig.~\ref{calibra}). The adopted SSP age is
12~Gyr. The last 2 lines report the calibration of 47 Tuc, using the
original data by Covino et al.~(1995) and those obtained by us on
their spectra by adopting the W94 index definition (see 3.2,
Table~1). In Col. 2 the empirical metallicities of the GCs as given
by Harris~(1996) are listed.}

\begin{tabular}{lcccccc}
\hline 
name &  \feh$_{\mathrm{ZW}}$& \multicolumn{2}{c}{$\mathrm{[Z/H]}^{SSP}_{WFFs}$} &
  \multicolumn{2}{c} {$\mathrm{[Z/H]}^{SSP}_{BFFs}$} & {$\mathrm{[Z/H]}^{SSP}_{BorFFs}$} \\ 
& & & & & \\ & & \mg  & \fe & \mg & \fe & \mg \\ 
\hline
 NGC~6981 & -1.40 & -1.91 & -1.83 & -1.87 & -1.73  & -1.46 \\ 
 NGC~6637 & -0.71 & -0.73 & -1.05 & -0.76 & -0.94  & -0.54 \\
 NGC~6356 & -0.50 & -0.51 & -0.91 & -0.55 & -0.81  & -0.35 \\ 
 NGC~6284 & -1.32 & -1.27 & -1.58 & -1.29 & -1.41  & -1.07 \\ 
 NGC~6626 & -1.45 & -1.31 & -1.48 & -1.33 & -1.31  & -1.12 \\ 
 NGC~6441 & -0.53 & -0.53 & -0.80 & -0.57 & -0.70  & -0.36 \\ 
 NGC~6218 & -1.48 & -1.80 & -1.71 & -1.77 & -1.57  & -1.35 \\ 
 NGC~6624 & -0.56 & -0.45 & -0.88 & -0.59 & -0.78  & -0.39 \\ 
 NGC~6388 & -0.60 & -0.81 & -0.84 & -0.84 & -0.74  & -0.61 \\ 
 NGC~5927 & -0.37 & -0.17 & -0.65 & -0.23 & -0.56  & -0.07 \\ 
 NGC~6553 & -0.34 & +0.05 & -0.27 & +0.01 & -0.29  & +0.11 \\
 NGC~6528 & -0.17 & +0.09 & -0.35 & +0.06 & -0.35  & +0.14 \\ 
47~Tuc(B84) & -0.76 & -0.50 & -0.66 & -0.53 & -0.58 & -0.33 \\ 
47~Tuc (W94) & -0.76 & -0.78  & -1.07 & -0.81 & -0.96 & -0.59  \\
\hline
\end{tabular}
\end{table*}

\subsection[]{The other Lick indices as abundance indicators}
\label{calibraother}
In this section we check the behaviour of the other Lick indices as
abundance indicators, by analysing the correlations with \mgb~and \fe,
which we have shown (Fig.~1) are likely to trace the $\alpha$~elements
and the Fe abundances, respectively. To this purpose, we divide the
indices into three groups: those which should be predominantly
sensitive to $\alpha$-elements (\mguno,\mg, etc.); those which should
trace the Fe abundance (Fe4383; Fe5782; etc.); and the others (\CNone;
\CNtwo; etc.), for which the case is less clear.  The correlations are
checked with respect to the model predictions. The evolutionary
populations synthesis of Maraston~(1998, see Sect.~3) has been updated
for the computations of the whole set of Lick indices given in Worthey
et al.~(1994), plus the higher-order Balmer lines of Worthey \&
Ottaviani~(1997). A detailed analysis of the contributions to these
indices, similar to that given in Sect.~4 for \mg~and \fe, will appear
in Maraston~(2003).
\begin{figure*} 
 \psfig{figure=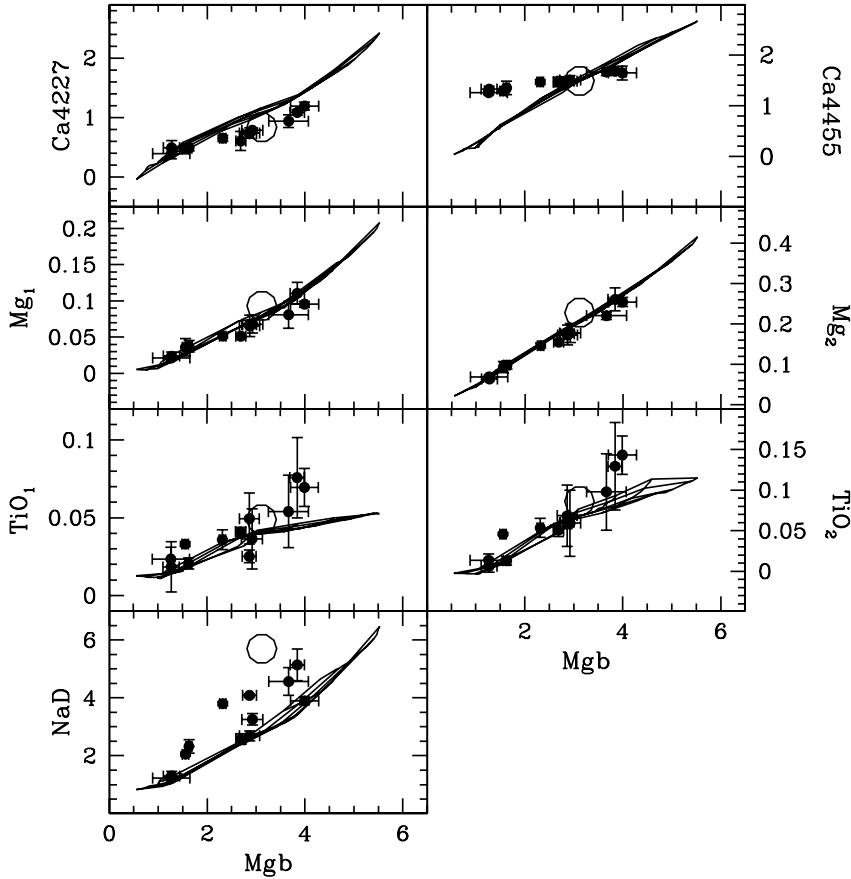,width=0.7\linewidth}
\caption{Calibration of $\alpha$-sensitive Lick indices. Standard SSPs
like in Fig.~1. Filled symbols denote our sample GCs, the large open
symbol the average value of the Bulge light in the Baade window.}
\label{alfa}
\end{figure*}

Fig.~\ref{alfa} shows the correlation of the first group of indices
with \mgb.  The Mg indices: \mgb,~\mguno~and \mg~are very well
consistent with each other and can be used as tracers of
$\alpha$-elements. The line-strength Ca4227 still appears to correlate
with Mgb, although displaying slightly smaller values, while the other
supposed Calcium-sensitive line Ca4455 does not. The reason for such
mismatch is not clear to us. We suspect a calibration problem. The
titanium-oxide indices \TiOone~and \TiOtwo~behave consistently with
\mgb~at low metallicities, while at high metallicities these indices
would overestimate the $\alpha$-element abundance with respect to
\mgb. However, TiO is contributed by M-type stars which may be poorly
treated in the models and which are present only in the metal-rich
clusters.
\begin{figure*}[!ht] 
 \psfig{figure=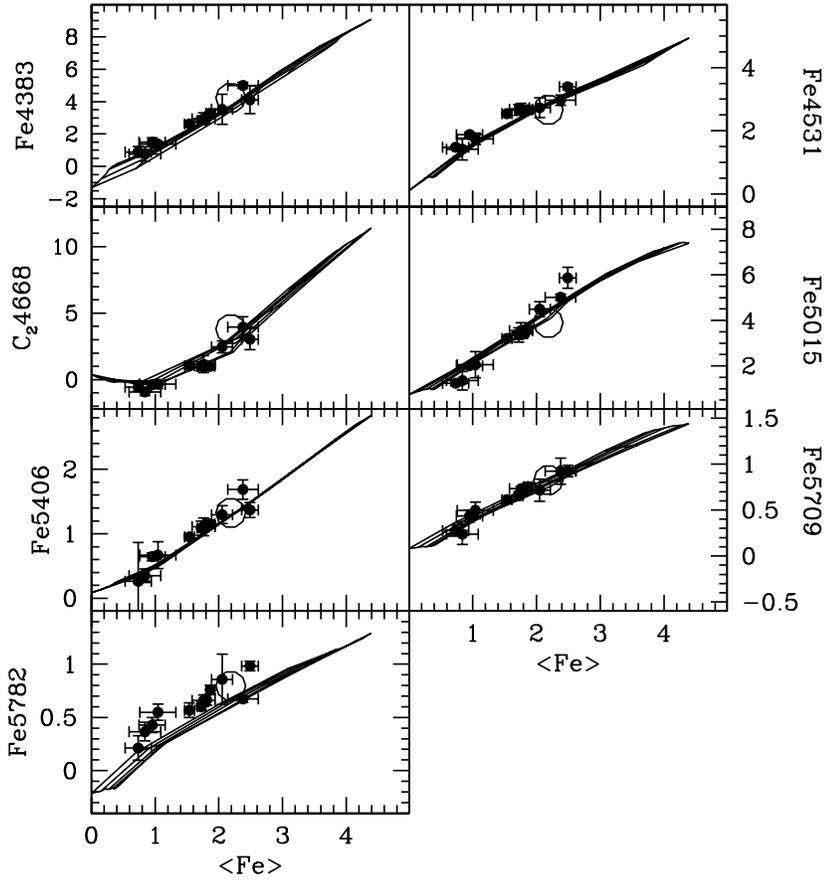,width=0.7\linewidth}
\caption{Model calibration. Iron-sensitive Lick indices. Model and data like in Fig.~\ref{alfa}.}
\label{ferri}
\end{figure*}
The NaD index scales with \mgb~although with some scatter. As
discussed by TMB02, the reason for the scatter is most probably the
absorption by interstellar medium affecting objects close to the
galactic plane. Because of this possible source of contamination the
NaD index is a problematic metallicity indicator for stellar
populations (see also Burstein et al.~1984).

Fig.~\ref{ferri} shows the second group of indices that should trace
the iron abundance. Indeed, all the plotted indices correlate very
well with \fe.  The models fit well most of these indices. The only
exception is Fe5782, which is underpredicted by the models, suggesting
an offset in the FFs. This offset most probably originates from the
low signal-to-noise of the Lick/IDS data for Fe5782, which results
from the extreme weakness of this line (S. Trager, {\it private
communication}).

It is worth noting that the \ferrofake~index is tighly correlated to \fe,
although is supposed to trace the abundance of Carbon (see Trager et
al.~1988).
\begin{figure*} 
 \psfig{figure=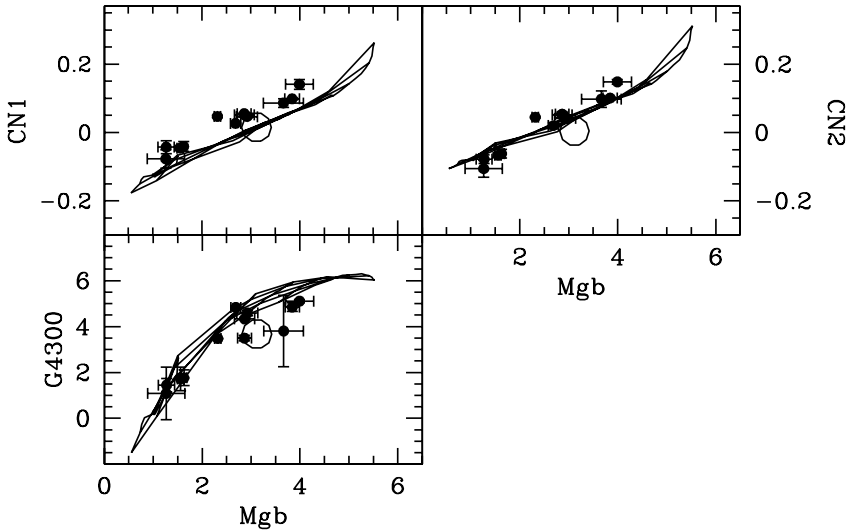,width=0.7\linewidth}
\caption{Model calibration of remaining Lick indices. Model and data
like in Fig.~\ref{alfa}.}
\label{altri}
\end{figure*}

In Fig.~\ref{altri} we plot the remaining indices vs \mgb.  \CNone~and
\CNtwo~seem to trace the same elements as \mgb, though with an offset,
especially at high metallicity and in the \CNone. At least part of the
effect could be due to the fitting functions not incorporating stars
belonging to high-metallicity GCs. As also discussed in Paper~I, GCs
have stronger CN indices compared to the bulge field (in Baade's
Window) at the same value of \fe. This may be caused by GCs stars
accreting the CN-rich ejecta of AGB stars. The low-density environment
of the field prevents a similar accretion on field stars. Therefore
the models at high metallicity are lower than GC data because the
fitting functions at high metallicity are contructed with field stars.

Finally also the G-band seems to follow magnesium, but the relation is more
scattered.

In Fig.s \ref{alfa} to \ref{altri}, the large open symbol shows the
average index values of 15 Bulge fields, located in Baade Window. In
all indices, the value of the Bulge field is consistent with an
average metallicity close to that of the most metal-rich GCs. This is
quantitatively confirmed by the detailed metallicity distribution of
the bulge stars from optical-infrared color-magnitude diagrams
(Zoccali et al.~2002). The large value of the NaD index for the Bulge
average field is again most probably due to contamination by
interstellar medium, as discussed above.

\subsection[]{Balmer lines}
\label{calibrabalmer}

\begin{figure} 
 \psfig{figure=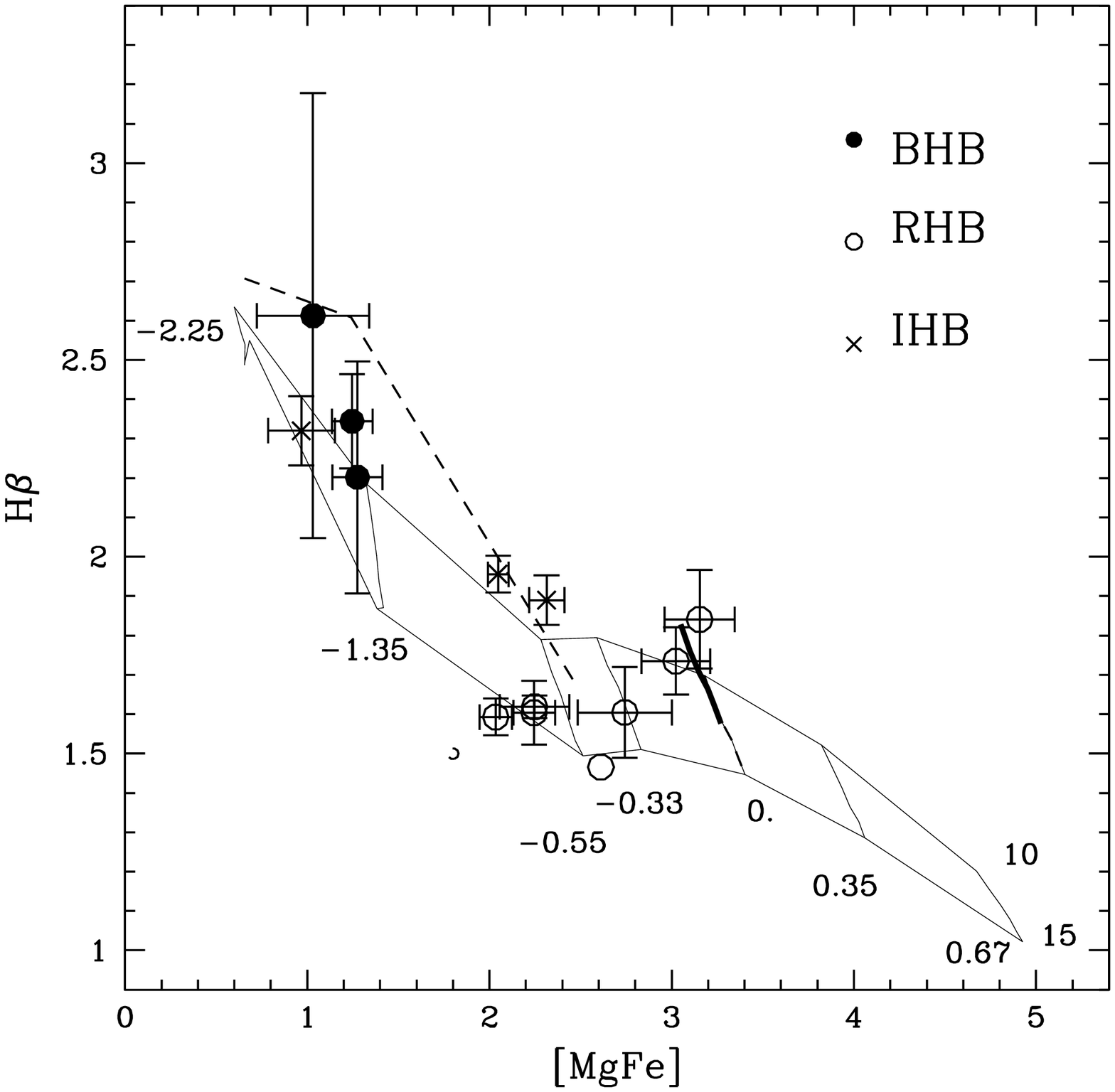,width=\linewidth}
\caption{Impact of the morphology of the Horizontal Branch on
\hb. Models refer to SSP ages of 10 and 15 Gyr, and $\zh$~from
$-2.25$~to $0.67$ as labeled in the Fig.. Solid lines show models in
which no mass loss has been accounted along the RGB. The dashed lines
are 15 Gyr old models in which mass-loss has been applied at
metallicities $\lapprox -0.5$, in order to reproduce the observed
\hb~of metal poor GCs (see Maraston \& Thomas~2000). The GCs of our
sample are plotted according to their observed HB morphology, by means
of the HBR parameter (from the Harris~1996 catalogue): pure blue HB
(HBR=1, solid symbols); red HB (HBR=0, open symbols); intermediate
morphologies (crosses). The open circle without errorbars shows the
average value of 15 Bulge fields in Baade window. The very thick line
at solar metallicity shows the locus of the solar metallicity models,
as corrected to take into account the bluening of the tracks due to
$\alpha$-enhancement (see Fig.~\ref{sala}).}
\label{hbeta}
\end{figure}
As well known Balmer lines are strongest in A-type stars, and become
pregressively weaker for decreasing temperatures. In synthetic stellar
populations their strength is sensitive to the temperature of the main
sequence turnoff, hence to the age. Therefore the Balmer line
strengths are used to estimate the age of e.g. elliptical galaxies, in
an attempt at breaking the age/metallicity degeneracy (e.g.~Worthey et
al.~1992). However, turnoff stars are not the only potential
contributor to the strength of the Balmer lines: Horizontal Branch
(HB) stars may be as warm or even warmer than the turnoff. Actually,
the \hb~index is perhaps more sensitive to the temperature
distribution of the HB (the HB morphology) than to any other parameter
(Worthey~1992; Barbuy \& de Freitas Pacheco~1995; Greggio~1997;
Maraston \& Thomas~2000). Therefore, attempting to break the
age/metallicity degeneracy with Balmer lines indices one runs into the
age/HB morphology degeneracy.  In the modeling the HB morphology
cannot be derived from first principles, because of the r\^ole played
by mass-loss on shaping the HB morphology . Additionally, possible
dynamical effects (e.g. Fusi Pecci et al.~1993) may determine
anomalous HB morphologies at a similar total cluster metallicity (the
2nd parameter problem in Milky Way GCs). Therefore the effect of the
HB morphology to be used in SSP models needs to be calibrated on the
Balmer indices.

For our models this is done in Maraston \& Thomas~(2000), to which we
refer for more details on the model parameter. In that work the
mass-loss to be applied at every SSP metallicity was calibrated in
order to reproduce the \hb~line of the GC sample by Burstein et
al.~(1984), Covino et al.~(1995) and Trager et al.~(1998).  Here we
check if those calibrated models are able to reproduce the Balmer
lines measured for our sample sample.

Fig.~\ref{hbeta} shows two sets of SSP models obtained with
different prescriptions for the RGB mass loss, various metallicities
and two ages (10 and 15 Gyr). Aimimg at calibrating the model Balmer
lines with metallicity, we have chosen as $x$-axis the
index~[MgFe]~\footnote{[MgFe]$=\sqrt{\mgb\cdot\fe}$}. TMB02 show that
this index by washing out \afe~effects, is able to trace total
metallicity.

The solid lines connect models in which no mass loss is applied to the
RGB. In this case the morphology of the HB is red (i.e., all HB
lifetime is spent on the red side of the RR-Lyrae location at
$\log\teff\sim3.85$), except at the very low metallicity $\zh=-2.25$,
where the HB is spent at $\log\teff\gapprox3.85$ even when no
mass-loss is applied. The dashed line shows the 15 Gyr SSPs with
$\zh\lapprox-0.5$, in which mass-loss has been applied on the RGB
according to canonical prescriptions. This leads to extended HBs, with
blue morphologies (i.e. the whole HB lifetime is spent the blue side
of the RR-Lyrae location) at $\zh=-2.25$, and intermediate HB
morphologies at metallicities between $\zh=-1.35$ ($\sim84\%$ of the
HB lifetime is spent blueward the RR-Lyrae) and
$\zh=-0.55$~($\sim10\%$ of HB lifetime is spent blueward the
RR-Lyrae).

The cluster data are plotted according to their observed HB
morphologies, (B(lue)HB: filled symbols; R(ed)HB: open symbols;
I(ntermediate)HB: asterisks) by means of the HBR parameter
(Harris~1996).

The calibrated models by Maraston \& Thomas~(2000) are able to
reproduce the observed \hb~of our sample GCs. In particular they
reproduce the relatively strong \hb~($\sim 1.9$~\AA) of NGC~6388 and
NGC~6441, which are metal-rich clusters ($\feh_{\mathrm{ZW}}=-0.6;-0.53$)
with an extension of the HB to the blue (Rich et al.~1997). The
percentage of HB stars that is found blueward the RR Lyrae gap is:
$\sim 15\%$ for NGC~6388 (Zoccali et al.~ 2000b) and $\sim 13\%$ for
NGC~6441~(M. Zoccali, {\it private communication}). This is well
consistent with the HB evolutionary timescales of the Maraston \&
Thomas (2000) models as given above.

Concerning the most metal-rich objects NGC~6528 and NGC~6553, their
relatively strong \hb~cannot be ascribed to HB effects since both
clusters have a red Horizontal Branch (Ortolani et al.~1995). Part of
the effect can be explained in terms of $\alpha$-enhancement at high
metallicities, without invoking young ages which would be in
contradictions with CMD determinations (Ortolani et al.~1995). In
Sect.~5 we have shown (Fig.~\ref{sala}) that $\alpha$-enhanced tracks
are bluer than the corresponding solar-scaled ones. As a consequence
the \hb~ lines are higher by $0.13$~\AA~at solar metallicity. The
thick line in Fig.~\ref{hbeta} connects the SSP models as shifted by
this amount.  Note however the rather large errorbar on the \hb~of
NGC~6553.

The index of the average light of the Bulge is shown as an open symbol
without errorbars, and sits on the $\sim$~15 Gyr model with red
HB. The low \hb~line does not leave room for intermediate age stars in
our sampled Bulge fields.
\begin{figure} 
 \psfig{figure=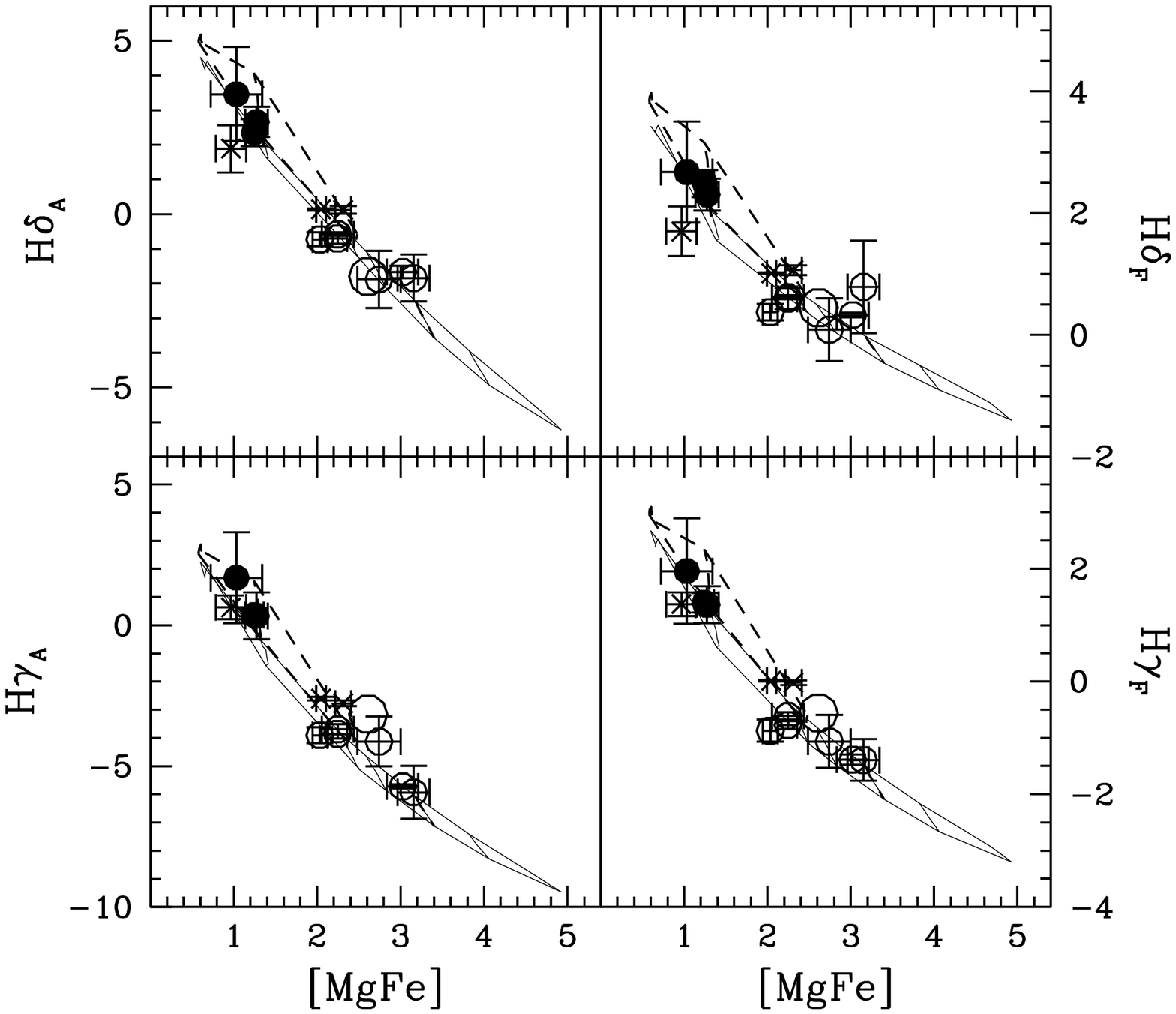,width=1.\linewidth}
\caption{Ca\-li\-bra\-tion of the hi\-gher-or\-der Bal\-mer li\-nes \hda,~\hdf,~\hga
and \hgf~(Wor\-they \& Ot\-ta\-via\-ni (1997) of the stan\-dard SSP mo\-dels, see
Fi\-gu\-re~\ref{hbeta}.}
\label{balmer}
\end{figure}

Finally we calibrate the higher-order Balmer lines
\hda,~\hga,~\hdf~and \hgf~(Worthey \& Ottaviani~(1997) of the same
models, in Fig.~\ref{balmer}. It can be appreciated a very good
consistency between the \hb and the higher-order Balmer lines.

\section[]{Summary and Conclusions}
\label{conclusion}

Synthetic Lick indices (e.g., \mg,~\fe,~\hb, etc.) of stellar
population models (SSP) are calibrated over a range of metallicities
that extends up to solar metallicity using a sample of galactic
globular clusters which includes high-metallicity clusters of the
Galactic bulge.  These data allow us to investigate {\it empirically}
a well known property of elliptical galaxies, known as ``magnesium
overabundance'' (Worthey et al.~1992), where the observed Mg indices
in ellipticals are much stronger at given iron than predicted by {\it
standard} models.  By standard models we mean those constructed using
stellar templates that are $\alpha$-enhanced at low metallicity but
assume solar elemental proportions at high metallicity.  This effect
has been generally interpreted in terms of $\alpha$-enhancement of
elliptical galaxies, even if the bulk of their stellar population is
metal rich.  However, such conclusion rests on two assumptions: i) the
models that are believed to represent solar scaled elemental ratios
are correct, and ii) the Lick Mg and Fe indices trace the abundance of
the corresponding element.

Using our GC database we have checked {\it empirically} both
assumptions by comparing the Lick indices of the GCs spectra with
those of ellipticals. The result is that the magnesium and iron
indices of the metal-rich GCs, of the integrated light of the Galactic
bulge, and of elliptical galaxies, define a fairly tight correlation
in the \mgb-\fe diagram, with elliptical galaxies lying on the
prolongation of the correlation established by the GCs, i.e., also the
metal-rich GCs of the Milky Way bulge exhibit the ``magnesium
overabundance'' syndrome.  Since the GCs are indeed known to have
enhanced \afe~ratios from stellar spectroscopy ($\afe\simeq +0.3$), we
conclude that the interpretation of elliptical galaxy spectra in terms
of ``magnesium overabundance'' is indeed correct. The comparison with
the GCs further allows us to point out that, rather than ``magnesium
overabundance'', the enhanced \afe~ratio in GCs, and ellipticals is
most probably the result of an {\it iron deficiency with respect to
the solar values (see Sect. 6.1). This agrees with previous
suggestions (Buzzoni et al.~1994; Trager et al. 2000b) and with the
recent results by TMB02}.  The enhanced \afe~ratio implies short
formation timescales for the bulk stellar population, an important
constraint for formation models of elliptical galaxies and bulges
(e.g., Matteucci~1994; Thomas et al.~1999), which is
difficult to reconcile with current semianalytic models of galaxy
formation (Thomas \& Kauffmann 1999).

In parallel, the comparison of the SSP model with the GC data has
allowed us to shed some light on the models themselves.  Around solar
metallicity, the standard models are based on the stellar indices of
Milky Way {\it disk} stars (by Worthey et al.~1994 or Buzzoni et
al.~1992; 1994), and therefore would reproduce the indices of stellar
populations with {\it solar-scaled} elemental proportions. This
explains why they fail to reproduce the \mgb-\fe correlation followed
by galactic GCs and the Galactic bulge, which are characterized by
elemental ratios which are specific to the Galactic {\it spheroid}.
This failure was also noted by Cohen et al.~(1998) for both the metal
rich globulars of the Milky Way as well as for those of M87, but was
not attributed to an abundance effect.

At metallicities $\zh\lapprox - 1.$ the standard models use metal
poor template stars that belong to the galactic halo, hence have
supersolar \afe~ratios. This explain why standard models successfully
reproduce the Lick indices of the metal poor GCs, which also belong to
the halo and are $\alpha$-element enhanced. The conclusion is that the
standard SSP models reflect abundance ratios which vary with
metallicity. This clearly complicates their use as abundance
indicators for extra-galactic stellar systems.

We have then proceeded to compare the data to a new set of SSP models
in which the \afe~ratio is treated as an independent variable (Thomas
et al.~2002b). The result is that the Galactic GCs and bulge, as well
as most ellipticals, are very well reproduced by coeval, old (12 Gyr)
models with $\afe=+0.3$, and various metallicities.  The uniqueness of
this $\alpha$-enhancement solution for the stellar populations of GCs,
the bulge, and ellipticals was checked by thoroughly exploring the
parameter space of the SSP models.  We find that the Lick indices are
little affected by the choice of the specific set of stellar
evolutionary tracks or fitting functions.  The only viable alternative
to abundance effects, which can produce high values of the Mg indices
coupled with low values of the Fe indices is a very steep IMF (much
steeper than Salpeter).  This solution, though formally acceptable, is
practically ruled out by many other observational constraints for the
clusters, as well as for the bulge and elliptical galaxies.

A closer look to the \mgb-\fe diagram reveals that elliptical
galaxies, unlike GCs, span a range of $\afe$~values, from just
marginally super-solar, to $\sim +0.4$.  Since the Mg index correlates
with the galaxy luminosity, the trend is in the direction of an
increasing $\alpha$-element overabundance with increasing luminosity
(mass). Apparently, the more massive the galaxy, the shorter the
duration of the star formation process (Thomas et al.~2002a). The
origin of this trend remains to be understood.

Our database of GCs was further used to check in an empirical fashion
the effectiveness of the other Lick indices to trace the element
abundances. Good indicators of $\alpha$-elements are found to be all
the Mg lines (\mg,~\mguno~and~\mgb), and \TiOone~and \TiOtwo~at
subsolar metallicities. Also the index Ca4227 does correlate with the
Mg indices, though a small offset between the two might be present.
Nearly all iron line indices (Fe4384, Fe4531, Fe5015, Fe5270, Fe5335)
are found to display very tight relations against another. The indices
\CNone,~\CNtwo~and the G-band G4300 follow Mg. On the contrary,
indices such as Ca4455, NaD and Fe5782 appear to be poorly calibrated,
and we cannot recommend their use as abundance indicators for
extra-galactic systems.

The Balmer \hb~plus the higher-order lines by Worthey \&
Ottaviani~(1997) \hda, \hdf, \hga, \& \hdf~are very well reproduced by
the standard SSP models considered here (Maraston \& Thomas, 2000;
Maraston~2003), which indicates that they are only marginally affected
by the \afe~ratio (see Tripicco \& Bell~1985; TMB02). Much more
important for their correct modeling is to account for the Horizontal
Branch morphology.  In particular the rather high Balmer lines
measured for NGC~6388 and NGC~6441 are modelled with a tail of warm
Horizontal Branch stars ($\sim~10\%$ of the total HB population).
These warm stars are indeed observed in the CMD of these two clusters
(Rich et al.~1997), in a number ($\sim 10\%$, Zoccali et al.~2000b)
which is in perfect agreement with the value required to reproduce the
strength of the Balmer lines.

Finally, we point out that the Mg indices of very metal poor stellar
populations ($\feh\sim -1.8$) are dominated by the contribution of the
lower main sequence. Therefore, these indices are prone to be affected
by the IMF and in GCs by the subsequent evolution of the mass function
due to the dynamical evolution of the clusters themselves.  It follows
that the Mg indices of very metal-poor stellar populations are not
reliable metallicity indicators.

\begin{acknowledgements}

We would like to thank Daniel Thomas, Beatriz Barbuy and Manuela
Zoccali for many enlightening discussions and Stefano Covino for
having provided the spectra of the globular clusters in his sample.
CM acwnowledges interesting discussions on Lick matters with Scott
Trager. We thank the referee Marcella Carollo for the prompt report
and the very positive comments. CM acknowledges the support by the
"Sonderforschungsbereich 375-95 f\"ur Astro-Teilchenphysik" of the
Deutsche Forschungsgemeinschaft.  LG acknowledges the hospitality of
Universit\"ats-Sternwarte M\"unchen where this research was carried
out.

\end{acknowledgements}

\end{document}